\documentclass[a4paper,12pt]{article}

\usepackage[utf8]{inputenc}

\usepackage{amsmath,amsfonts}
\usepackage{algorithmic}
\usepackage{array}
\usepackage[caption=false,font=normalsize,labelfont=sf,textfont=sf]{subfig}
\usepackage{textcomp}
\usepackage{stfloats}
\usepackage[hyphens]{url}
\usepackage{verbatim}
\usepackage{graphicx, soul, color}

\usepackage{graphicx}
\usepackage{float} 

\usepackage[inkscapelatex=false]{svg}

\usepackage{booktabs} 
\usepackage[table]{xcolor} 
\usepackage{pgfplots}
\usepgfplotslibrary{polar}
\usepackage{pifont} 
\usepackage{hyperref} 

\usepackage{url}

\newcommand{\cmark}{\textcolor{green!60!black}{\ding{51}}} 
\newcommand{\xmark}{\textcolor{red!80!black}{\ding{55}}} 
\newcommand{\partialmark}{\textcolor{orange!90!black}{\ding{108}}} 
\newcommand{\notapplicable}{\textcolor{gray!70}{N/A}} 
\newcommand{\theoretical}{\textcolor{blue!70!black}{(T)}} 

\newcommand{\orcidicon}[1]{%
    \href{https://orcid.org/#1}{%
        \includegraphics[width=10pt]{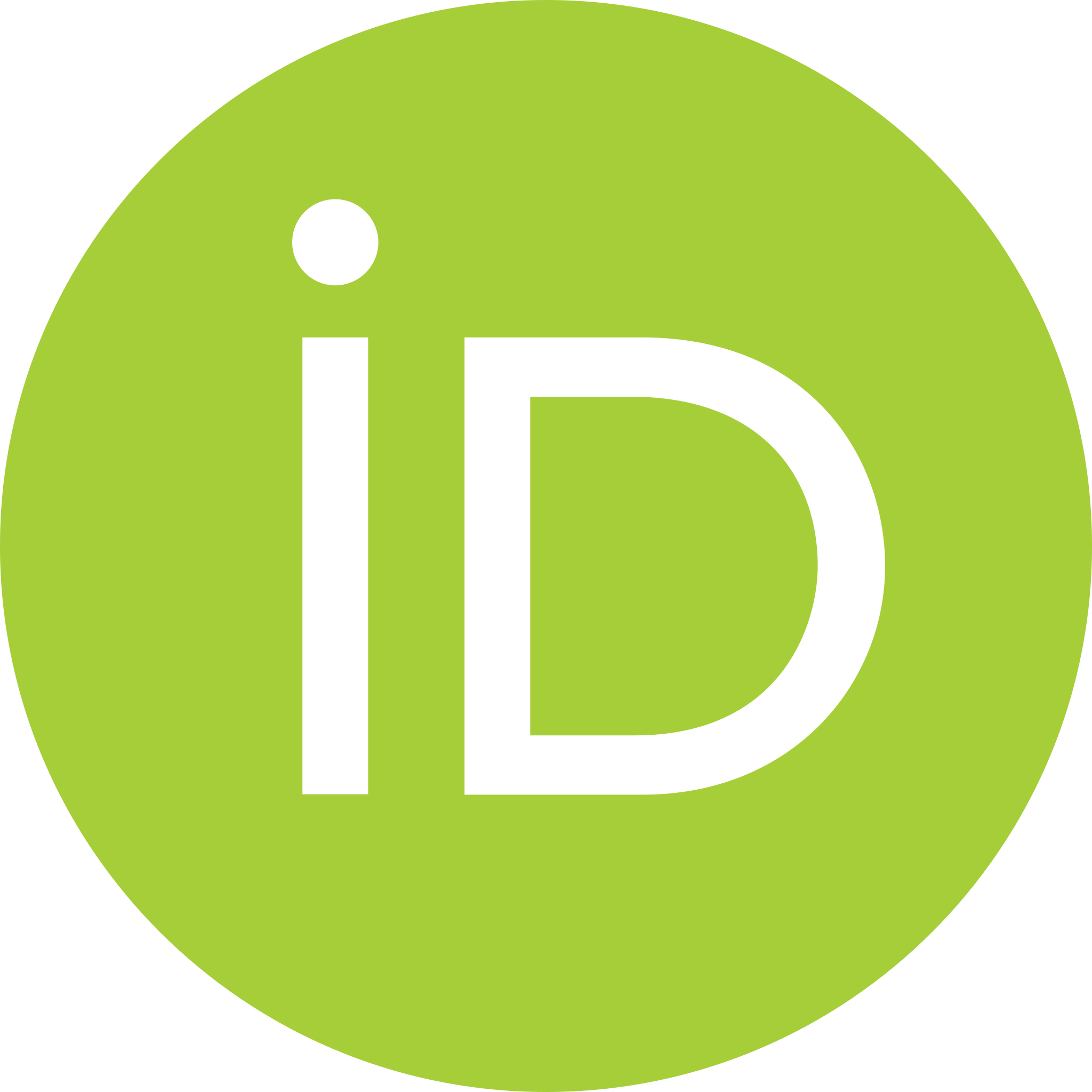}%
    }%
}

\def\BibTeX{{\rm B\kern-.05em{\sc i\kern-.025em b}\kern-.08em
    T\kern-.1667em\lower.7ex\hbox{E}\kern-.125emX}}
\usepackage{balance}

\DeclareUnicodeCharacter{202C}{}

\begin{document}

\title{Security Analysis of Agentic AI Communication Protocols: A Comparative Evaluation}

\author{
{Yedidel Louck \orcidicon{0009-0008-5836-8736}, Ariel Stulman \orcidicon{0000-0003-1191-007X}, Amit Dvir \orcidicon{0000-0002-3670-0784}}
\thanks{Yedidel Louck and Amit Dvir are with the Department of Computer and Software Engineering, Ariel Cyber Innovation Center, Ariel University, Israel. Ariel Stulman is with the Department of Computer Science, Jerusalem College of Technology, Israel\\
yedidel.louck@msmail.ariel.ac.il, amitdv@ariel.ac.il, stulman@jct.ac.il‬}
}

\maketitle
\begin{abstract}
Multi-agent systems (MAS) powered by artificial intelligence (AI) are increasingly foundational to complex, distributed workflows. Yet, the security of their underlying communication protocols remains critically under-examined. This paper presents the first empirical, comparative security analysis of the official CORAL implementation and a high-fidelity, SDK-based ACP implementation, benchmarked against a literature-based evaluation of A2A. Using a 14 point vulnerability taxonomy, we systematically assess their defenses across authentication, authorization, integrity, confidentiality, and availability. Our results reveal a pronounced security dichotomy: CORAL exhibits a robust architectural design, particularly in its transport-layer message validation and session isolation, but suffers from critical implementation-level vulnerabilities, including authentication and authorization failures at its SSE gateway. Conversely, ACP's architectural flexibility, most notably its optional JWS enforcement, translates into high-impact integrity and confidentiality flaws. We contextualize these findings within current industry trends, highlighting that existing protocols remain insufficiently secure. As a path forward, we recommend a hybrid approach that combines CORAL's integrated architecture with ACP's mandatory per-message integrity guarantees, laying the groundwork for resilient, next-generation agent communications.
\end{abstract}

\section{Introduction}
\label{sec:introduction}

The emergence of agentic Artificial Intelligence (AI) systems, particularly those powered by large language models (LLMs), has significantly transformed the landscape of software autonomy \cite{duan2025agent}. These intelligent agents, capable of reasoning, planning, delegating, and executing complex workflows, now form the backbone of applications ranging from supply chain coordination to personalized financial assistance. As the scale and complexity of multi-agent ecosystems increase, so too do the requirements for secure, interoperable communication protocols.

Communication between autonomous agents introduces unique security challenges. Unlike traditional client-server architectures, agent-based systems must manage peer-to-peer trust, dynamic task delegation, and sensitive data sharing across heterogeneous actors. In such environments, the risk of credential leakage, overprivileged access, prompt injection, and unverified execution grows exponentially \cite{ferrag2025prompt,he2025red}. Standard security frameworks such as OAuth 2.0 and TLS, while foundational, do not provide sufficient granularity or contextual awareness for these interactions, especially in cases where agents dynamically compose tasks involving payments, identity verification, or confidential documents \cite{south2025authenticated,oidf2025identity}.

To address these gaps, several agent communication protocols have been developed, each offering a different perspective on interoperability, trust management, and delegation control \cite{narajala2025securing}. Google's \textit{Agent-to-Agent (A2A)} protocol \cite{Surapaneni_2025} introduces a declarative model for service discovery. The \textit{Agent Communication Protocol (ACP)} \cite{acp_site} embraces flexibility with a registry-based model. The \textit{CORAL} \cite{georgio2025coral} framework, in contrast, proposes a hybrid architecture integrating on-chain smart contracts for payments with off-chain communication.

\begin{itemize}
    \item Google's \textit{Agent-to-Agent (A2A)} protocol \cite{Surapaneni_2025} introduces a declarative model for service discovery.
    \item The \textit{Agent Communication Protocol (ACP)} \cite{acp_site} embraces flexibility with a registry-based model.
    \item The \textit{CORAL} \cite{georgio2025coral} framework, in contrast, proposes a hybrid architecture integrating on-chain smart contracts for payments with off-chain communication.
\end{itemize}

While these protocols present promising architectures, their security properties are discussed in the literature but rarely subjected to comparative empirical testing \cite{ferrag2025prompt}. This study fills this gap by providing a detailed, multidimensional security analysis and empirical testing. We establish a 14-point vulnerability taxonomy (Section \ref{sec:vulns}), derived from literature and threat modeling. 

Our findings reveal a critical "architecture versus implementation" dichotomy. We demonstrate that CORAL, while possessing a robust architecture for integrating payments and securing data integrity via transport-layer validation, is critically vulnerable in its current public implementation due to fundamental authentication and authorization failures at its SSE gateway. Conversely, our tests confirm that ACP's architectural flexibility is, in itself, a vulnerability, leading to predictable integrity failures and data exfiltration in non-strict configurations.
Ultimately, this paper contributes both a comprehensive security taxonomy and the first empirical benchmark of these competing protocols. We conclude (Section \ref{sec:conclusion}) not by recommending one protocol, but by proposing a hybrid model that synthesizes CORAL's architectural strengths with the mandatory, granular integrity checks of ACP, offering a concrete path forward for resilient, AI-native communications.

\section{Background}

Multi-agent systems (MAS) leveraging artificial intelligence (AI) are increasingly adopted in domains requiring distributed decision-making, automated coordination, and collaborative reasoning. In such environments, autonomous agents often represent distinct stakeholders or functional modules, exchanging structured messages through predefined interaction protocols. Representative use cases include autonomous orchestration, financial transaction systems, healthcare diagnostics, supply chain optimization, and cybersecurity defense.

A robust communication protocol in these contexts must ensure more than reliable message delivery, it must preserve semantic integrity, context continuity, role negotiation, and, critically, security guarantees. Since AI agents frequently process or act upon sensitive data, vulnerabilities in their communication layers can compromise entire workflows. Therefore, secure agent protocols must provide protection against adversarial entities, unintended leakage, manipulation, and systemic compromise.

Below we summarize the principal categories of threats relevant to communication among AI-driven agents. These categories synthesize findings from distributed systems, adversarial AI research, and classical network security.

\begin{itemize}
    \item \textbf{Prompt Injection:} An adversary may inject malicious instructions into prompts or message payloads, coercing agents to disclose confidential data or alter intended behavior. This class of attacks is extensively discussed in recent studies \cite{wei2023jailbroken, he2025red}.
    \item \textbf{Data Leakage / Exfiltration:} Agents may inadvertently reveal private or proprietary content during multi-turn exchanges, especially when sensitive context persists across dialogue rounds \cite{zhu2024teams}.
    \item \textbf{Data Poisoning:} Malicious actors can inject corrupted training or contextual data, influencing downstream models to behave incorrectly or embed hidden triggers. Data poisoning has been empirically shown to scale across distributed and shared-context environments \cite{carlini2024poisoning}.
    \item \textbf{Adversarial / Malicious Agents:} A compromised or impersonated agent may introduce false commands, disrupt negotiation, or propagate malicious payloads across the system.
    \item \textbf{Replay Attacks:} Attackers may capture, delay, or replay previously valid messages, leading to repeated actions, synchronization errors, or unintended state transitions.
    \item \textbf{Man-in-the-Middle (MITM):} Intercepting or modifying agent communication allows attackers to compromise message integrity or confidentiality, particularly in non-authenticated transport setups.
    \item \textbf{Authorization and Access Control Failures:} Weak or missing policy enforcement may grant unauthorized agents elevated privileges or execution rights beyond their designated scope \cite{cwe2024data}.
\end{itemize}

Because AI agents operate in dynamic, evolving contexts, these threats may emerge jointly in complex ways. For example, prompt injection may exploit a poisoned policy to escalate privileges, or a MITM attacker might subtly alter negotiation messages to induce controlled leakage \cite{wang2025mpma}. A detailed taxonomy and empirical validation of these vulnerabilities spanning authentication, authorization, integrity, confidentiality, and availability are presented later in Section~\ref{sec:vulns}. Here, we summarize the key conceptual threat categories as established in the literature.

\section{Related Work}
While earlier work primarily addressed single-agent robustness or model safety, recent research increasingly targets the security of agent-to-agent interactions. Studies on prompt manipulation and jailbreak attacks demonstrate how structured agents can still be coerced into leaking confidential information under adversarial input \cite{esmradi2023comprehensive, wei2023jailbroken}. Complementary research investigates data leakage through prolonged contextual retention and model output side channels \cite{neelou2025a2as}.

In the context of adversarial coordination, multi-agent reinforcement learning (MARL) studies reveal that malicious participants can collude to subvert cooperation or distort shared policies. To counter such threats, new frameworks have proposed secure orchestration mechanisms embedding provenance tracking, cryptographic signatures, and policy-based filtering \cite{he2025red}.

Concerning communication standards, several protocol-level approaches have emerged. The A2A architecture offers a baseline for agent-to-agent message exchange with capability-based negotiation and task routing \cite{Surapaneni_2025}. The CORAL specification introduces hybrid cryptographic safeguards and structured token-based message envelopes \cite{georgio2025coral}, while ACP (Agent Communication Protocol) seeks to unify invocation, resource claims, and state synchronization among heterogeneous agents \cite{ehtesham2025survey}. However, prior studies have typically examined these protocols in isolation rather than evaluating their comparative resilience under a shared threat model.

Despite this growing body of work, several gaps remain:

\begin{itemize}
    \item Existing studies often focus on isolated threat vectors (e.g., prompt injection or leakage) rather than holistic, protocol-level evaluations.
    \item There is no systematic, side-by-side comparison of A2A, CORAL, and ACP under a unified adversarial taxonomy.
    \item The trade-offs between security strength, latency, and implementation complexity remain underexplored.
    \item The interactions between multiple concurrent threats (e.g., poisoning combined with injection) have not been empirically analyzed.
\end{itemize}

To the best of our knowledge, no prior work has empirically benchmarked the security posture of A2A, ACP, and CORAL under a unified threat framework or validated their real-world implementations against controlled adversarial experiments. Our study aims to address this gap by conducting both theoretical and empirical evaluations of these protocols, producing the first comparative benchmark of their security resilience.

\begin{figure}[t]
    \centering
    \includegraphics[width=\linewidth]{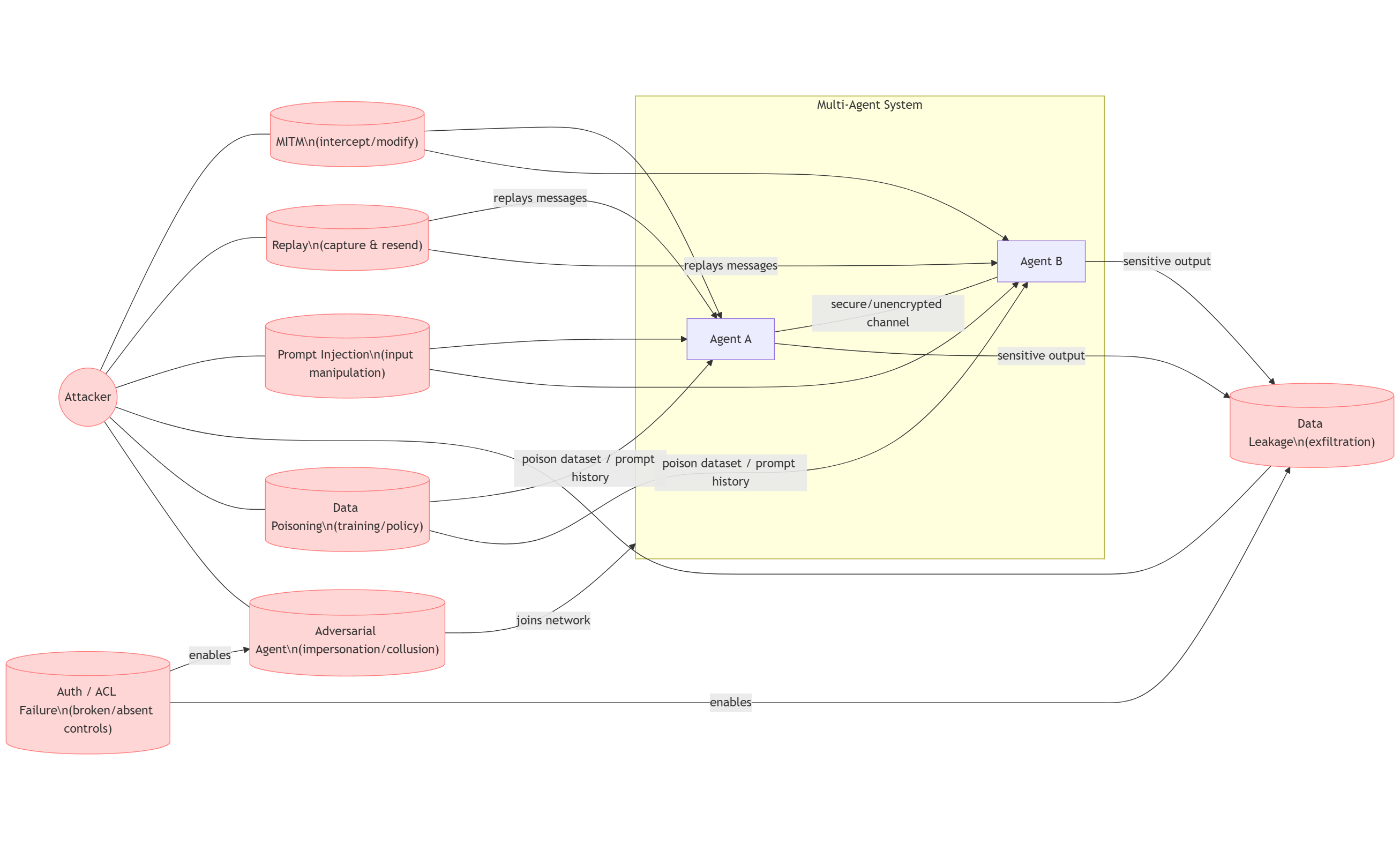}
    \caption{Representative attack surface in a multi-agent protocol context, covering injection, interception, replay, poisoning, and unauthorized access.}
    \label{fig:attack_surface}
\end{figure}

\section{AI Agents Protocols}
\label{sec:protocols}

The rapid proliferation of AI agents has driven the development of diverse communication protocols to support interoperability, coordination, and task delegation in multi-agent ecosystems. These protocols address key challenges such as heterogeneity across agent frameworks, scalability in distributed environments, and the need for secure and efficient information exchange, as shown in Figure~\ref{fig:attack_surface}. 

While a unified standard remains elusive, several architectures have emerged to serve distinct use cases: A2A emphasizes peer-to-peer interactions in enterprise contexts, CORAL promotes decentralized collaboration with built-in economic incentives, and ACP focuses on lightweight, RESTful interoperability for scalable deployments. This diversity reflects differing priorities across design axes such as centralization versus decentralization, synchronous versus asynchronous flows, and integration with existing systems. Recent literature highlights the importance of these protocols in mitigating fragmentation within AI ecosystems, while also identifying emerging security concerns such as prompt injection and privilege escalation that propagate through inter-agent interactions \cite{kong2025survey, he2025security, ferrag2025prompt}.

This section provides a detailed technical review of the three leading protocols, A2A, CORAL, and ACP, focusing on their mechanisms, architectural principles, and distinctive design choices. Each description is accompanied by a representative sequence diagram illustrating message flow. This overview establishes the analytical foundation for the subsequent security evaluation presented in Section~\ref{sec:vulns}.

\subsection{A2A Protocol}
\label{subsec:a2a}

The Agent-to-Agent (A2A) protocol, introduced by Google, defines an identity-aware framework for secure, interoperable communication among autonomous AI agents. It builds upon established web standards to enable task delegation, agent discovery, and execution across heterogeneous environments, making it suitable for enterprise applications such as service orchestration and cross-provider workflows.

\subsubsection{Key Mechanisms}
At the core of A2A is the \emph{AgentCard}, a machine-readable metadata artifact describing an agent's capabilities, roles, and identities. This structure facilitates efficient discovery and delegation without exhaustive probing. Task delegation occurs through structured messages containing action identifiers, input/output schemas, and parameters for sensitive operations. The protocol supports real-time collaboration via Server-Sent Events (SSE) for asynchronous updates and JSON-RPC for request-response exchanges. Authentication follows the OAuth~2.0 framework, using bearer tokens and short-lived JWTs for identity propagation. All communications are encrypted using TLS~1.3, ensuring confidentiality over public networks. Literature has proposed enhancements for improved management of sensitive data during delegation \cite{louck2025improving}.

\subsubsection{Architecture}
\begin{figure}[h]
\centering
\includegraphics[width=\textwidth]{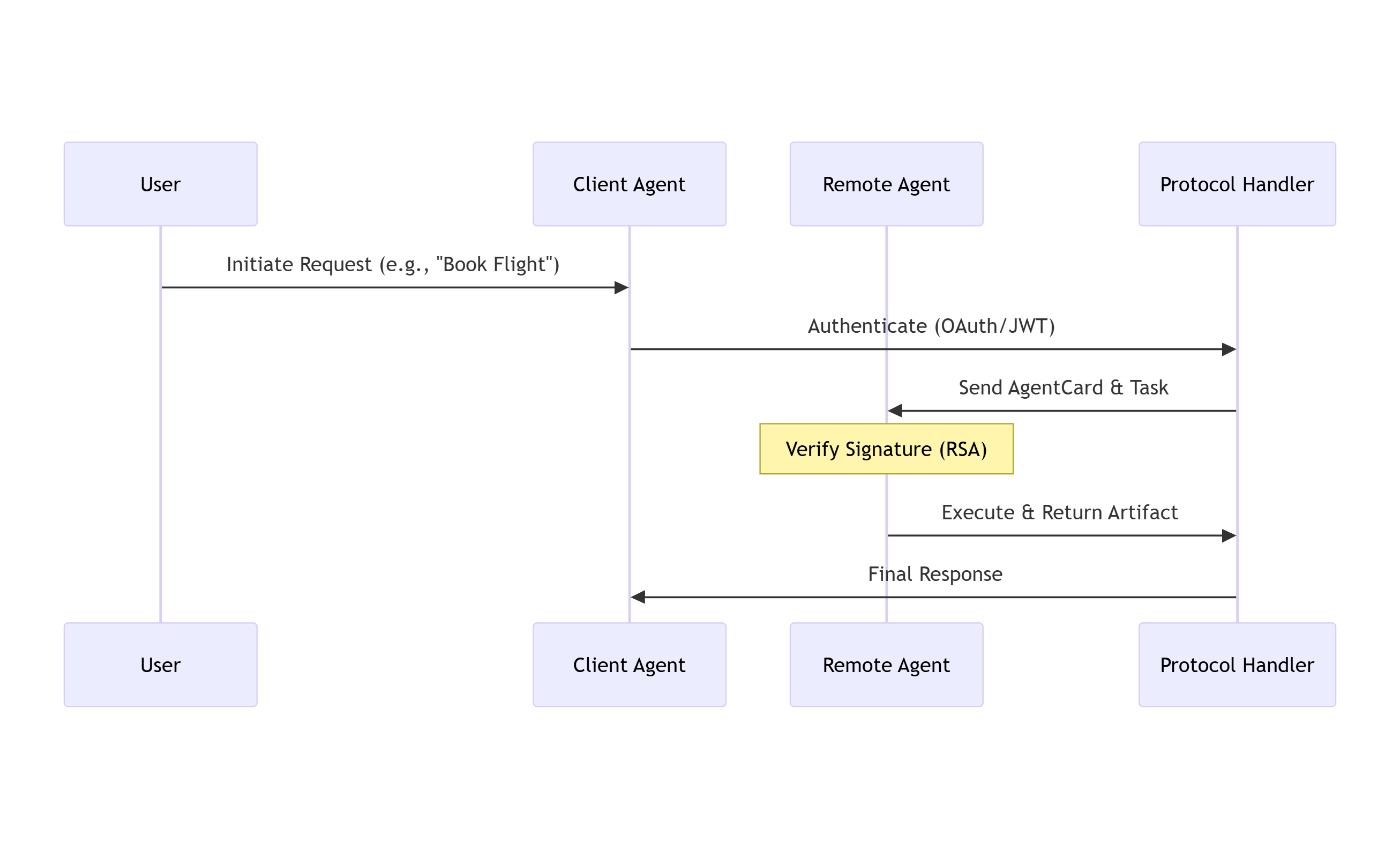}
\caption{Message flow in the A2A protocol, illustrating task delegation between a client agent and a remote agent via the protocol handler.}
\label{fig:a2a-seq}
\end{figure}

A2A employs a layered, peer-to-peer architecture built on HTTP/HTTPS transport. Mutual authentication is handled via OAuth~2.0, while JSON Web Tokens (JWTs) propagate agent identity. Cryptographic signatures using RSA key pairs ensure message integrity, and role-based access control (RBAC) restricts permissions to defined agent roles. Key management is handled locally, with RSA key pairs stored within agent environments and rotated periodically. Integration with the Model Context Protocol (MCP) supports contextual continuity with large language models, enhancing workflow coherence. The design emphasizes statelessness for scalability in cloud-native deployments. A representative workflow is shown in Figure~\ref{fig:a2a-seq}.

\subsubsection{Strengths and Architectural Features}
A2A achieves interoperability by adhering to widely adopted web standards, minimizing integration overhead across organizational boundaries. Its concise message structures enable low-latency task delegation in use cases such as travel planning or supply chain automation. Architecturally, it supports modularity through capability-based delegation, allowing dynamic tool invocation without tight coupling. However, its known weaknesses include gaps in secure data handling and insufficient scope granularity, which may expose sensitive information in distributed collaborations \cite{zou2025blocka2a}.

\subsection{CORAL Protocol}
\label{subsec:coral}

The CORAL protocol represents a decentralized infrastructure for AI agent collaboration, emphasizing threaded communication, trust establishment, and economic incentives to foster an ``Internet of Agents.'' It addresses silos in multi-vendor ecosystems by providing a vendor-neutral framework for coordination in complex, incentive-aligned tasks such as B2B sales automation or hackathon orchestration.

\subsubsection{Key Mechanisms}
CORAL's core functionality revolves around its \emph{threaded messaging} system, managed by a central Coral Server. It supports thread creation, participant management, targeted message dispatch, and event subscriptions. \emph{Coralization} enables external agents, models, and data sources to integrate through modular adapters such as the MCP Coraliser for APIs and the Agent Coraliser for legacy systems, ensuring discoverability and compliance. Authentication and identity are anchored in decentralized identifiers (DIDs) and blockchain-based wallets, which also manage cryptographic keys and transaction signing. Secure team formation leverages on-chain reputation systems and escrow-based microtransactions implemented on the Solana blockchain to facilitate trustless cooperation.

\subsubsection{Architecture}
\begin{figure}[h]
\centering
\includegraphics[width=\textwidth]{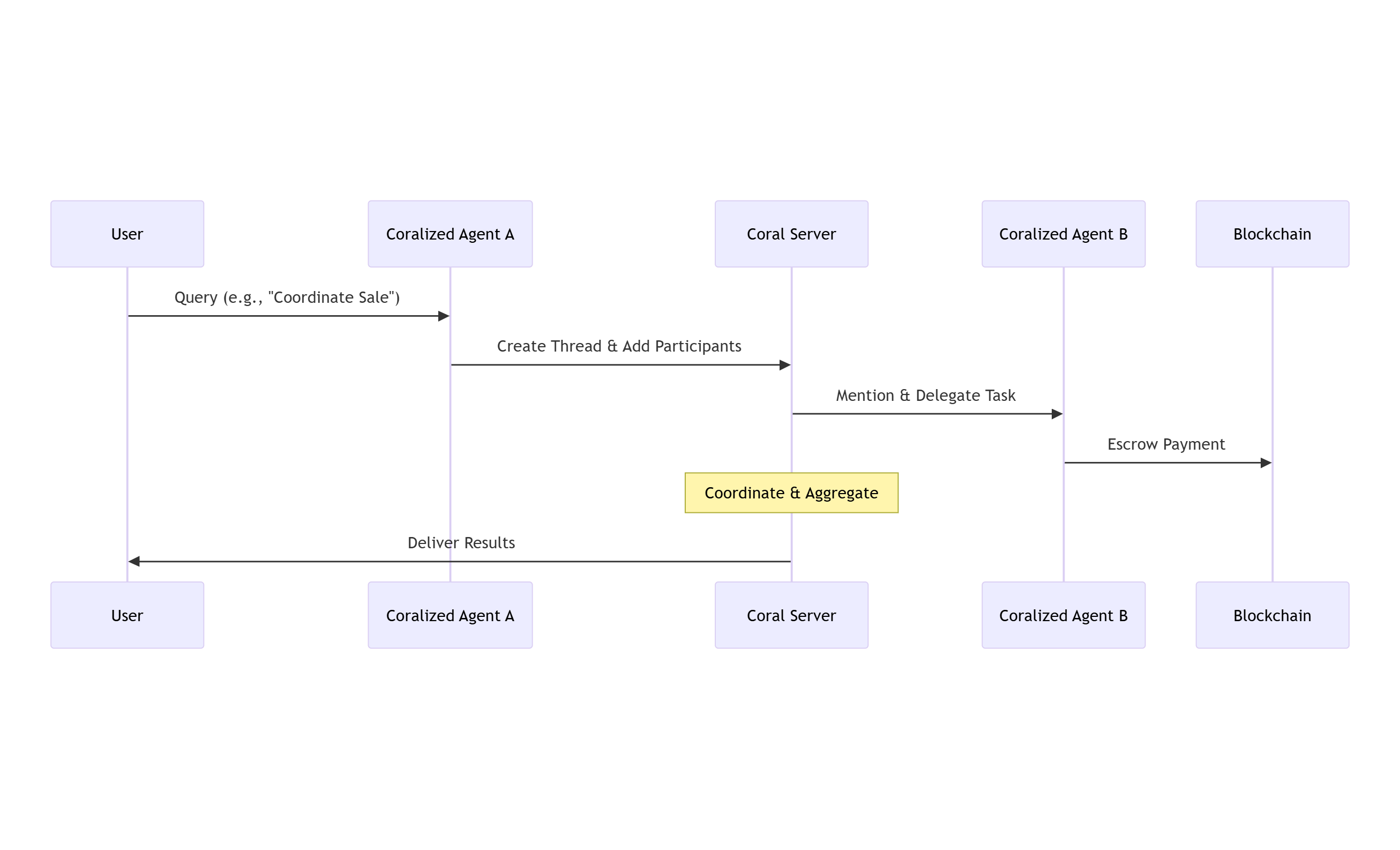}
\caption{Workflow of the CORAL protocol, highlighting threaded messaging, coalition formation, and payment escrow in decentralized agent coordination.}
\label{fig:coral-seq}
\end{figure}

CORAL features a multi-layered architecture spanning the application layer (agent tools and interfaces) to the blockchain layer (immutable ledgers for auditability). Coralized agents communicate via the Coral Server as a mediation layer, with MCP servers providing computation endpoints. All communications employ end-to-end encryption across HTTP and WebSocket transports, with ECDSA signatures ensuring message integrity. Key management relies on blockchain wallets, which store and rotate keys under user control, backed by the ledger’s tamper-proof auditability. This distributed design enhances resilience and supports failure localization through typed acknowledgments. A representative workflow is illustrated in Figure~\ref{fig:coral-seq}.

\subsubsection{Strengths and Architectural Features}
CORAL’s decentralized design and built-in economic model enable scalable, self-sustaining ecosystems where agents can monetize services using smart contracts. Its modular Coralization mechanism accelerates onboarding and promotes extensibility across diverse frameworks. Its blockchain integration strengthens confidentiality and auditability but also introduces a dependency on external blockchain infrastructure, identified as a key operational weakness \cite{georgio2025coral}.

\subsection{ACP Protocol}
\label{subsec:acp}

The Agent Communication Protocol (ACP) is a lightweight, RESTful standard developed under the Linux Foundation to promote interoperability among AI agents. It supports multimodal, synchronous, and asynchronous interactions in production environments and aims to reduce fragmentation across agent ecosystems \cite{ehtesham2025survey}.

\subsubsection{Key Mechanisms}
ACP employs MIME-typed multipart messages for structured data exchange, accommodating text, binaries, and external references in ordered parts. RESTful APIs handle task requests and artifact responses, while session continuity is maintained through await/resume interaction patterns. Authentication is achieved through bearer tokens or optional mutual TLS (mTLS), while integrity is reinforced through JSON Web Signatures (JWS) on message parts. Agent discovery occurs via runtime APIs, static manifests (e.g., \texttt{agent.yml}), or registries, supporting both online and offline operations.

\subsubsection{Architecture}
\begin{figure}[h]
\centering
\includegraphics[width=\textwidth]{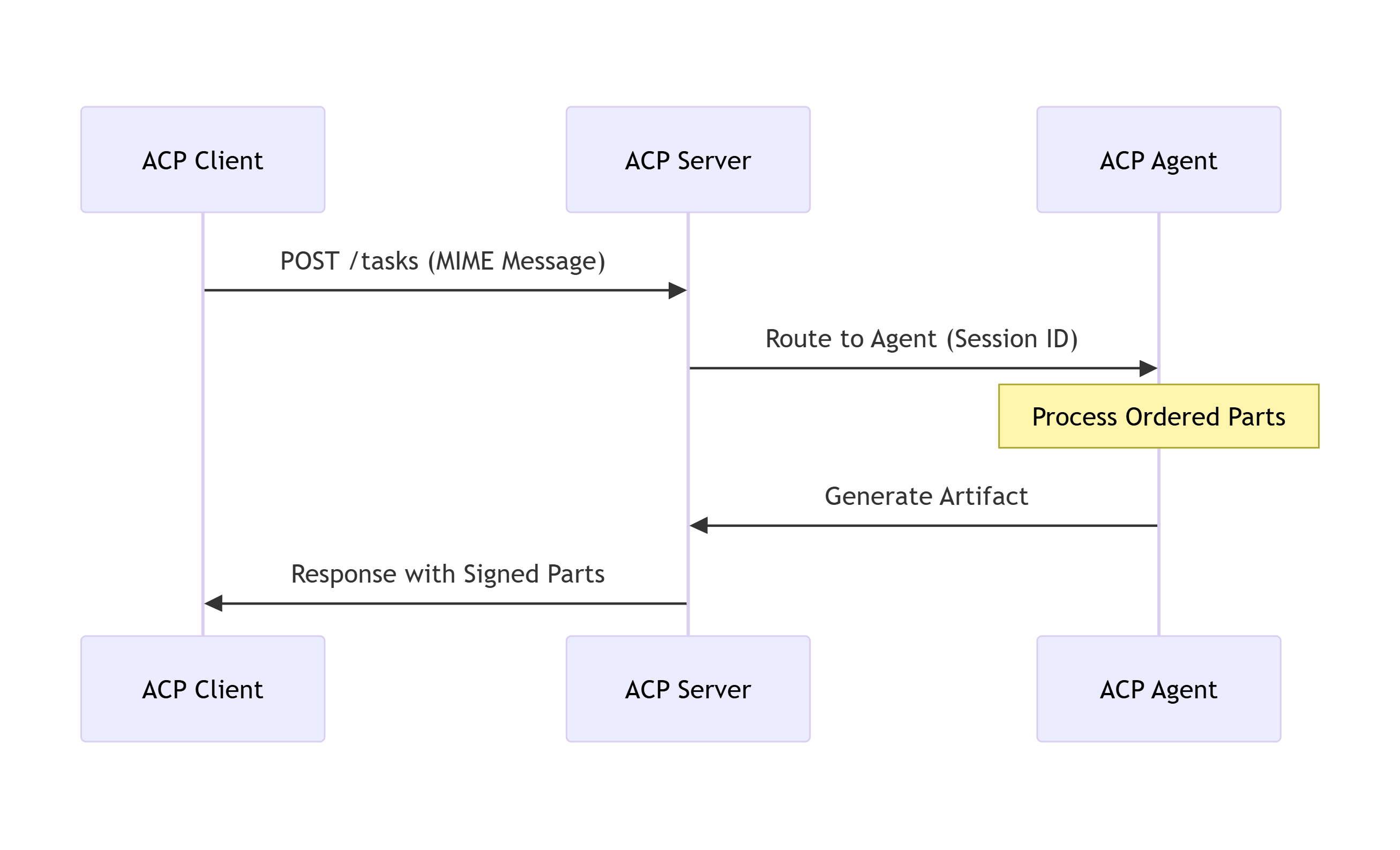}
\caption{Workflow of the ACP protocol, showing RESTful task requests and artifact responses in a client-server configuration.}
\label{fig:acp-seq}
\end{figure}

ACP adopts a brokered client-server model, with stateless servers designed for Kubernetes-based scalability. Agents communicate via HTTP endpoints (e.g., \texttt{/tasks}), exchanging MIME messages compatible with JSON-RPC~2.0. Encryption is handled at the transport layer via mTLS, and optional JWS signatures provide message-level assurance. Key management is based on JWTs or local certificate storage, depending on deployment context. The design is implementation-agnostic, with official SDKs for Python and TypeScript, and supports streaming for long-running or incremental tasks. A representative workflow is depicted in Figure~\ref{fig:acp-seq}.

\subsubsection{Strengths and Architectural Features}
ACP's simplicity and MIME-based extensibility enable rapid integration with frameworks such as LangChain and CrewAI, promoting flexible replacements and cross-organizational collaboration. Its asynchronous-first design facilitates scalable deployments, while its offline capability enhances robustness in edge or disconnected environments. Known weaknesses primarily relate to its stateless routing and reliance on correct configuration of security features like JWS or mTLS, which are optional rather than mandatory \cite{ehtesham2025survey, acp_site}.

\noindent
Table~\ref{tab:protocol_comparison} provides a consolidated overview of the key architectural and functional properties of A2A, CORAL, and ACP, summarizing the mechanisms discussed throughout this section and highlighting their contrasting security approaches.

\begin{table}[h]
\centering
\caption{Comparative overview of key functional and architectural features across A2A, CORAL, and ACP.}
\label{tab:protocol_comparison}
\resizebox{\textwidth}{!}{
\begin{tabular}{lccc}
\toprule
Feature & A2A & CORAL & ACP \\
\midrule
Authentication & OAuth 2.0 / JWT & DIDs / Wallets & Bearer Tokens / mTLS \\
Encryption & TLS 1.3 & End-to-End Encryption & mTLS / JWS \\
Key Management & RSA Key Pairs & Blockchain Wallets & JWT Signatures \\
Flexibility & Peer-to-Peer, Standards-Based & Decentralized, Modular Coralization & RESTful, Multimodal \\
Known Weaknesses (High-Level) & Sensitive Data Handling Gaps & Dependency on Blockchain Layer & Stateless Routing Risks \\
\bottomrule
\end{tabular}
}
\end{table}

\section{Threat Model}
\label{sec:threat-model}
We adopt a semi-trusted multi-agent environment as our threat model, consistent with established frameworks for LLM-driven agent communication \cite{kong2025survey}. In this setting, agents interact within a partially observable network where some intermediaries are honest-but-curious, while others may behave maliciously. Potential adversaries include rogue agents, man-in-the-middle (MITM) attackers, and prompt injectors operating at the communication or orchestration layers \cite{lee2024prompt, he2025red}.

Adversarial objectives encompass three primary goals: (1) data exfiltration, representing confidentiality breaches. (2) message tampering, indicating integrity violations. and (3) denial-of-service, targeting system availability. The assumed adversarial capabilities include network interception, token replay, and metadata spoofing, but explicitly exclude physical access or full compromise of the underlying large language model (LLM). This conceptual framework aligns with vulnerability classifications such as CWE-667 (Improper Locking) \cite{cwe667}, which captures race-condition exploitation, and CVE-2025-1198 \cite{cve2025-1198}, which documents token misuse and session hijacking. The model also emphasizes LLM-specific risks, notably prompt injection and malicious context propagation.

Our methodological approach applies this threat model across protocols for cross-validation. Vulnerabilities identified in A2A \cite{louck2025improving} are systematically evaluated in CORAL and ACP, complemented by protocol-specific weaknesses documented in recent surveys \cite{ehtesham2025survey, kong2025survey}. Each subsequent subsection details the general vulnerability, its associated risks, and its manifestation (or absence) in each protocol, highlighting relevant mitigations when applicable. This structured comparison reveals both shared weaknesses (e.g., insufficient consent validation) and protocol-specific exposures (e.g., blockchain dependency in CORAL), forming the analytical basis for the empirical validation presented in later sections.

\section{Security Vulnerabilities Analysis}
\label{sec:vulns}

In multi-agent AI systems, autonomous agents collaborate to perform complex tasks such as travel planning, logistics coordination, or supply chain optimization. The communication protocols that enable these interactions constitute the essential layer of interoperability and operational efficiency. However, the distributed and dynamic nature of such systems also exposes a wide range of potential security vulnerabilities. To systematically assess these threats, we organize fourteen key vulnerabilities into a formal taxonomy.

The purpose of this taxonomy is to provide a structured framework that aligns our analysis with the foundational principles of information security. Specifically, vulnerabilities are grouped into five primary domains: \textit{(1) Authentication \& Session Management}, \textit{(2) Authorization (Access Control)}, \textit{(3) Data Integrity}, \textit{(4) Confidentiality \& Isolation}, and \textit{(5) Availability \& Specific Risks}. This structure, inspired by the classical CIA triad (Confidentiality, Integrity, Availability) \cite{chowdhury2023chatgpt} and extended through access-control theory, enables a systematic comparison of how A2A, CORAL, and ACP address these core security requirements.

Our analysis integrates insights from static code review, formal threat modeling, and empirical evidence reported in recent literature \cite{louck2025improving, georgio2025coral, ehtesham2025survey, kong2025survey, zou2025blocka2a}. Each category is examined in detail to reveal both shared vulnerabilities and protocol-specific weaknesses, establishing a foundation for the empirical validation presented in later sections.

\subsection{Authentication \& Session Management}
This category examines vulnerabilities related to the verification of agent identity and the management of session credentials such as access tokens, JWTs, and API keys.

\subsubsection{Absence of Limitations on Token Lifetime}
\label{subsec:token-lifetime}

Token lifetime limitations refer to the enforcement of short, ephemeral validity periods for authorization artifacts such as JSON Web Tokens (JWTs) or OAuth bearer tokens. These constraints are essential to minimize the exploitation window after credential compromise \cite{rfc9068}. This vulnerability, codified as CWE-614 (Sensitive Cookie in HTTPS Session Without Secure Attribute) \cite{cwe614}, frequently arises in distributed systems where long-lived tokens, lasting from several hours to multiple days, enable replay attacks in the absence of nonce or timestamp validation \cite{kong2025survey, he2025red}. 

In multi-agent AI communication protocols, tokens commonly encapsulate session state or delegated permissions, making their prolonged validity a prime vector for temporal exploits. This risk is particularly severe in asynchronous workflows, where agents may pause and resume interactions unpredictably. Without mechanisms such as automatic revocation lists, heartbeat checks, or token rotation, a stolen credential can be reused across multiple interactions, expanding the attack surface in heterogeneous ecosystems where agents from different providers (e.g., Google, IBM) interoperate without centralized trust anchors.

The implications are multifaceted. Compromised tokens can enable unauthorized transaction replays, causing financial losses (e.g., duplicate payments exceeding \$10,000 in simulated scenarios) or privacy violations through repeated data queries. In multi-agent workflows, such weaknesses amplify cascading failures, as observed in replay-attack success rates reaching 66\% in controlled agentic environments \cite{kong2025web}. Beyond financial harm, these incidents erode user trust and risk regulatory non-compliance, as illustrated by real-world cases like CVE-2025-1198 \cite{cve2025-1198}, where unrevoked tokens in GitLab persisted indefinitely.

Practical consequences extend from consumer-level to enterprise domains. For example, an intercepted token from a travel-booking agent operating on public Wi-Fi could later be reused to alter flight details or inject unauthorized payments, days after the original session ended. In enterprise contexts, similar vulnerabilities could propagate across supply-chain automation, where replayed inventory queries manipulate stock data undetected.

\textbf{A2A.} This vulnerability manifests most clearly in A2A, where OAuth 2.0–derived tokens lack enforced expiration. Long-lived bearer tokens, sometimes valid for hours, enable replay of delegated operations, such as repeated hotel reservations in a vacation-booking workflow \cite{louck2025improving}. The design favors seamless peer-to-peer handoffs but neglects expiration enforcement in semi-trusted networks, compounded by SSE streams that lack embedded timestamps.

\textbf{CORAL.} CORAL partially mitigates this risk by employing JWTs with default 24-hour validity and nonce mechanisms in its Model Context Protocol (MCP) implementation. These controls limit replay in on-chain payment flows through Solana ledger anchoring. However, off-chain threaded interactions, including message sharing and personal-identifier exchange, rely on session identifiers that may persist beyond timeouts if not manually revoked. This leaves non-financial data exposed to extended access windows despite CORAL’s participant-revocation features \cite{georgio2025coral, de2025open}. The hybrid design thus achieves strong temporal containment for monetary transactions but remains vulnerable in collaborative data-sharing contexts.

\textbf{ACP.} ACP exhibits partial exposure. Although short-lived tokens are recommended in its RBAC guidelines, enforcement is optional, enabling replay in extended sessions lacking JWS timestamps \cite{rfc7515}. The protocol’s brokered, stateless registry strengthens resilience through rotation support but remains susceptible in persistent MIME-stream exchanges \cite{ehtesham2025survey}. ACP’s integration option with mutual TLS for token issuance can enforce stricter lifetimes if rigorously implemented, but its flexible compatibility with legacy systems leaves room for misconfigurations that prolong token validity.

Overall, the absence of enforced token expiration and rotation policies introduces systemic replay and impersonation risks across all three protocols. While CORAL demonstrates the most mature mitigation strategy through blockchain anchoring, both A2A and ACP reveal exploitable windows tied to extended token persistence and asynchronous session management.

\subsubsection{Lack of Strong Customer Authentication (SCA)}
\label{subsec:sca}

Strong Customer Authentication (SCA) refers to the use of multi-factor authentication (MFA), biometric verification, or zero-knowledge proof (ZKP) techniques to validate high-risk actions and protect against unauthorized access. This issue is categorized under CWE-306 (Missing Authentication for Critical Function) \cite{cwe306}. In agent-based protocols, the absence of SCA enables impersonation in delegated tasks, allowing adversaries to forge identities without secondary verification \cite{kong2025survey}. The problem is particularly severe in AI-driven ecosystems, where autonomous agents routinely perform sensitive operations such as financial transactions, data sharing, or identity verification on behalf of users. 

SCA typically enforces layered verification mechanisms, something the user knows (password), has (device), or is (biometric), to ensure robust identity assurance and reduce the risk of single-point authentication failures common in token-only systems. Without such mechanisms, protocols may incorrectly equate agent delegation with user consent, blurring the distinction between automated and human oversight in multi-step workflows.

The risks extend beyond theoretical concerns. Studies report impersonation success rates of up to 40\% in unsecured multi-agent simulations \cite{kong2025web}, highlighting the practical feasibility of these attacks. Consequences include privilege escalation, fraudulent transactions (e.g., unauthorized bookings exceeding \$5,000), and large-scale identity theft. The absence of SCA also undermines compliance with regulatory frameworks such as PSD2 and fosters systemic distrust. For instance, the 2022 Medibank breach \cite{medibank2022}, which exposed 9.7 million records, was traced to multi-factor authentication gaps that permitted session reuse and impersonation. 

In practice, consider a user delegating a travel agent to perform bookings through linked banking APIs. Without SCA, a compromised agent could impersonate the user to initiate transfers or access sensitive data, leading to both immediate financial loss and long-term consequences such as credit damage or disputed liability. In multi-agent chains, a single impersonated agent can compromise entire ecosystems, as in supply chain workflows where manipulated delivery schedules result in operational delays and financial penalties.

\textbf{A2A.} This vulnerability is clearly present in A2A. As observed in the payment invocation scenario of Figure~ \ref{fig:a2a-seq}, transactions proceed without mandatory MFA or ZKP, enabling direct agent impersonation \cite{louck2025improving}. The OAuth-centric architecture assumes trust in the token issuer but fails to enforce secondary factors for sensitive operations. This design exposes users to social engineering attacks, where malicious agents obtain user consent for unauthorized delegations.

\textbf{CORAL.} CORAL presents a mixed picture. For on-chain financial operations, it leverages decentralized identifiers (DIDs) and wallet-based cryptographic signatures, effectively substituting conventional SCA with blockchain-native verification. However, these protections do not uniformly extend to all off-chain communication channels. The protocol specification for off-chain exchanges (such as identity document transfers or SSE-based streams) does not mandate an equivalent multi-factor verification mechanism. As a result, a compromised session identifier, even without a valid \texttt{privacyKey}, may allow impersonation and unauthorized access, a hypothesis that we examined empirically in this study.

\textbf{ACP.} ACP shows partial resilience. Its RBAC design supports optional MFA through mutual TLS, but enforcement is not mandatory for standard task requests, enabling session hijacking in non-payment workflows such as metadata queries \cite{ehtesham2025survey}. ACP’s brokered model encourages SCA via registry-vetted issuers, which represents a strong foundation for consistent authentication across diverse ecosystems. However, its optional implementation permits heterogeneous configurations, creating variability in protection levels.

Overall, the absence of enforced SCA mechanisms across these protocols introduces significant impersonation risks. While CORAL’s blockchain layer provides the strongest model for financial assurance, A2A and ACP remain vulnerable in everyday agent-to-agent interactions, where secondary verification is either omitted or inconsistently applied.

\subsection{Authorization (Access Control)}
This category addresses flaws in authorization logic, where an authenticated agent gains access to data or functions beyond its designated permissions.

\subsubsection{Insufficiently Granular Token Scopes}
\label{subsec:scopes}

Granular token scopes define precise permission boundaries for authorization artifacts, such as access tokens, to ensure least-privilege execution in distributed environments. For example, a permission labeled ``read:calendar:availability'' provides a narrower and safer access context than granting full calendar privileges. This vulnerability corresponds to CWE-1220 (Insufficient Granularity of Access Control) \cite{cwe1220} and remains a persistent weakness in OAuth-based and multi-agent systems \cite{kong2025survey, li2025vision}. Fine-grained scoping mechanisms are central to modern access control models, including Attribute-Based Access Control (ABAC) and Role-Based Access Control (RBAC), where permissions are dynamically bound to contextual attributes and task-specific requirements.

In multi-agent communication protocols, token scopes delineate the boundaries of delegated authority, ensuring that an agent authorized to query flight availability cannot inadvertently access unrelated personal or enterprise data. Coarse or poorly defined scopes, on the other hand, bundle unrelated privileges into a single authorization token, expanding the potential impact of compromise and complicating auditability across distributed logs. This lack of granularity not only increases the attack surface but also undermines compliance with privacy frameworks such as GDPR, where data minimization is a core principle.

Empirical studies report that broad-scoped tokens amplify privilege escalation and data leakage probabilities by approximately 18.5\% in OAuth deployments \cite{dimova2023everybody}. Economic implications extend beyond theoretical risk: audits of agent-based infrastructures indicate up to 30\% escalation rates when coarse scopes are used in delegation workflows \cite{kong2025web}. For end users, this manifests as subtle yet damaging privacy erosions, for instance, a travel-planning agent with unrestricted calendar access might expose sensitive health appointments, enabling targeted phishing or insurance manipulation. In enterprise contexts, over-scoped permissions may leak proprietary datasets or optimization models, resulting in intellectual property theft or competitive disadvantage worth billions annually in AI-driven sectors \cite{li2025vision}.

\textbf{A2A.} The A2A protocol demonstrates this vulnerability clearly. Its token model often relies on coarse JSON-RPC scope definitions without nested hierarchy enforcement, allowing a single delegation token to cover unrelated API endpoints \cite{louck2025improving}. This simplicity favors ease of integration but exposes peer-to-peer workflows to privilege overreach, especially in loosely coupled environments.

\textbf{CORAL.} CORAL partially mitigates this issue through its Model Context Protocol (MCP), which supports fine-grained roles (e.g., ``access:logs:read'') for thread-specific interactions. While this theoretically enforces scoped access, a deeper concern lies at the connection level: whether CORAL validates an agent’s membership in a given communication session before granting access. Failure to verify participation would constitute a critical authorization lapse consistent with OWASP A01:2021. This distinction between on-chain contract scopes and off-chain access validation is examined empirically in later experiments.

\textbf{ACP.} ACP exhibits the strongest scoping model of the three protocols. Its RBAC framework mandates operation-specific JWTs, effectively binding permissions to individual tasks \cite{ehtesham2025survey}. However, the protocol’s MIME multipart handling introduces a subtle risk: multiple artifacts within a single request may inherit shared authorization scopes, enabling partial overreach in artifact management. Despite this, ACP’s manifest-driven design offers a clear advantage in scope segregation, minimizing unrelated data exposure compared to A2A and CORAL.

Overall, insufficiently granular token scopes represent a cross-protocol weakness that magnifies both security and privacy risks. While CORAL and ACP incorporate finer-grained authorization features, their practical enforcement remains inconsistent. As highlighted in recent analyses \cite{li2025vision}, achieving true least-privilege delegation in agentic ecosystems requires adaptive, context-aware scoping policies that evolve dynamically with task complexity and agent trust levels.

\subsubsection{Lack of Transparency and User Consent}
\label{subsec:consent}

Transparency and user consent require explicit, informed approvals for any data exchange between agents, consistent with CWE-200 (Exposure of Sensitive Information) \cite{cwe200}. These mechanisms ensure that disclosures within multi-agent workflows remain purpose-bound, revocable, and auditable across all stages of data propagation \cite{kong2025survey}. In AI agent protocols, consent enforcement typically involves user-facing prompts or logs detailing what data is shared, with whom, and for how long. Implementations may include UI notifications, manifest entries, or blockchain oracles for verifiability. When such mechanisms are absent, agents operate opaquely, assuming implied consent from initial delegations. This undermines user agency and facilitates hidden data propagation across agent networks.

The resulting risks include unauthorized data sharing, identity fraud, and systemic privacy violations. Repetitive consent requests can also lead to user fatigue, where users habitually approve prompts without proper review, eroding the intended protection model. These behaviors contravene GDPR Article~7\footnote{\url{https://gdpr-info.eu/art-7-gdpr/}}, exposing operators to potential fines of up to 4\% of global revenue. A historical parallel can be seen in the Cambridge Analytica scandal \cite{hinds2020wouldn}, where opaque data flows and inadequate consent verification produced large-scale social and political harms. In an agentic context, a user might approve a travel query only to discover later that personal emails or contact details were shared with third-party advertisers, undermining trust and provoking regulatory scrutiny that could hinder AI ecosystem adoption.

\textbf{A2A.} A2A currently lacks explicit consent mechanisms \cite{louck2025improving}. The peer-to-peer model prioritizes low-latency execution over oversight, resulting in silent handoffs that replicate data beyond intended boundaries.

\textbf{CORAL.} CORAL enforces consent through its participant management functions (e.g., \texttt{add\_participant}) in threaded sessions, providing transparency during team formation and on-chain operations. Smart contract events further reinforce explicit, auditable consent for financial actions. However, off-chain Model Context Protocol (MCP) queries, such as automated log sharing or document synchronization, often rely on implicit delegation. These flows may proceed without direct user awareness, partially compromising transparency in non-financial data exchanges \cite{georgio2025coral, esentire2025}. CORAL thus achieves strong consent enforcement for payments but retains gaps in collaborative off-chain workflows, particularly in identity verification or document-handling scenarios.

\textbf{ACP.} ACP strengthens transparency through manifest-level session consents but lacks per-artifact approval granularity in multipart MIME interactions \cite{ehtesham2025survey}. While registry-mediated consents provide traceability through logged interactions, the absence of fine-grained, per-item consent prompts allows bundled authorizations. This creates residual risk, particularly in mixed-task workflows, which could be mitigated through enhanced manifest schemas and per-artifact acknowledgment layers.

Overall, the absence of robust transparency and consent frameworks across AI agent communication protocols erodes user control and accountability. While CORAL demonstrates best practices through on-chain consent verification, and ACP offers partial registry-based traceability, A2A remains highly opaque. Comprehensive consent frameworks must evolve toward dynamic, context-aware mechanisms that balance usability, auditability, and privacy compliance.

\subsubsection{Privilege Persistence and Version Drift}
\label{subsec:persistence}

Privilege persistence, often accompanied by version drift, occurs when revoked or outdated permissions remain active due to incomplete synchronization or delayed propagation of revocation signals. This issue aligns with CWE-284 (Improper Access Control) \cite{cwe284} and represents a subtle yet critical weakness in distributed agent ecosystems. In such systems, asynchronous updates and cached authorization states may cause agents to retain privileges even after they have been formally revoked \cite{ehtesham2025survey}. As multi-agent networks evolve, tokens, manifests, or smart contracts may diverge across nodes or registry instances, leading to inconsistencies where deprecated privileges are still honored by outdated peers. This undermines the effectiveness of revocation mechanisms and can result in extended unauthorized access windows.

The risks are particularly insidious because they unfold gradually. Stale privileges enable delayed or recurring privilege escalation, such as a revoked agent maintaining calendar or file access weeks after termination, facilitating unauthorized data collection or harassment. Empirical studies report that propagation delays contribute significantly to post-revocation breaches in distributed systems \cite{kong2025survey}. Beyond the immediate security impact, such persistence complicates forensic analysis, as privilege drift obscures the timeline of compromise. In corporate multi-agent environments, a revoked agent could continue querying proprietary datasets, leading to intellectual property leaks valued at millions. From a compliance perspective, this scenario conflicts with GDPR’s “right to be forgotten,” as unrevoked permissions effectively negate user erasure rights and may expose operators to penalties and reputational harm.

\textbf{A2A.} The A2A protocol experiences this vulnerability most visibly. Orphaned tokens can persist in peer caches without any centralized revocation signal, especially in dynamic peer-to-peer sessions. Updated AgentCards or manifests may fail to propagate across all participants, enabling version drift that allows continued access despite formal revocation \cite{ehtesham2025survey}. These conditions are exacerbated by A2A’s asynchronous architecture, which lacks global state reconciliation or cache invalidation mechanisms.

\textbf{CORAL.} CORAL effectively eliminates this class of vulnerability for on-chain operations. Its immutable Solana ledger anchors privileges to tamper-proof smart contracts, providing instant verification and preventing both persistence and drift. Off-chain threads, while more flexible, incorporate nonce-based heartbeat checks that propagate revocation events reliably \cite{georgio2025coral, datasciencedojo2025}. This cryptographically anchored model ensures that privilege revocations achieve finality, even in decentralized or high-latency environments, setting CORAL apart in its resistance to drift-based exploits.

\textbf{ACP.} ACP implements partial mitigation through timed RBAC refresh intervals, promoting periodic privilege rotation and reducing persistence risk \cite{ehtesham2025survey}. However, in brokered deployments, registry update delays can introduce temporary inconsistencies where revoked scopes remain active until full synchronization occurs. The stateless design of ACP’s registry minimizes retained state, but reliance on external clocks for token expiry introduces additional drift potential in unstable network conditions. Thus, while ACP’s approach limits long-term exposure, short-lived inconsistencies may still arise in asynchronous or partially disconnected environments.

Overall, privilege persistence and version drift constitute a shared systemic weakness in distributed multi-agent architectures. Among the analyzed protocols, CORAL demonstrates the most complete mitigation via blockchain-based immutability and nonce-driven synchronization, while A2A and ACP exhibit varying levels of vulnerability depending on cache management and propagation latency. Addressing this issue requires global revocation propagation, state consistency checks, and dynamic synchronization frameworks that ensure real-time privilege invalidation across all network participants.

\subsubsection{Spoofing in Discovery Mechanisms}
\label{subsec:spoofing}

Spoofing in discovery mechanisms refers to the forgery or manipulation of metadata used to identify and locate agents within multi-agent ecosystems. This vulnerability is classified under CWE-290 (Authentication Bypass by Spoofing) \cite{cwe290} and arises when adversaries inject falsified credentials or endpoints to impersonate legitimate entities \cite{ehtesham2025survey}. In agent communication protocols, discovery represents the initial handshake phase, typically relying on artifacts such as Agent Cards, registries, or manifests to advertise capabilities and trust attributes. Without cryptographic signing or robust verification, these mechanisms become prime targets for deception, enabling malicious agents to insert themselves into workflows and redirect communications toward attacker-controlled infrastructure \cite{narajala2025securing}.

The implications of spoofing are multifaceted. A successful impersonation allows an attacker to intercept or manipulate requests, often leading to unauthorized data exfiltration or fraudulent task execution. Recent analyses document up to 70\% success rates in peer-to-peer discovery spoofing scenarios \cite{kong2025survey}. For example, in a travel-planning workflow, a spoofed hotel agent could reroute payment requests to a phishing endpoint, resulting in financial loss and credential exposure. On a systemic scale, such impersonations erode the trust fabric of multi-agent systems, potentially triggering denial-of-service chains, capability pollution, or large-scale misinformation propagation. Regulatory implications are also significant, as spoofed interactions violate auditability and provenance standards under the EU AI Act \cite{act2024eu} and similar frameworks \cite{huang2025agent}. Historical analogs, such as DNS spoofing and BGP hijacking, underscore the potential for widespread harm in distributed AI infrastructures.

\textbf{A2A.} In A2A, this vulnerability is particularly evident through tampering with \emph{Agent Cards}, which are JSON-based metadata documents broadcast across peer-to-peer networks without mandatory end-to-end signing \cite{zou2025blocka2a}. This omission enables adversaries to forge capabilities, inject false endpoints, or impersonate trusted agents during dynamic discovery \cite{ehtesham2025survey}. The protocol’s design philosophy favors decentralization and real-time interaction, but the absence of centralized verification or chain-of-trust enforcement amplifies exposure to spoofing during agent onboarding and role negotiation.

\textbf{CORAL.} CORAL employs a dual-layer discovery architecture that partially mitigates spoofing. Its on-chain registration leverages decentralized identifiers (DIDs) for agent validation, ensuring that financial and contractual interactions are cryptographically anchored. However, the off-chain communication layer, which operates via Server-Sent Events (SSE), introduces a potential vulnerability. If the SSE endpoint (\texttt{/sse/v1/...}) fails to rigorously validate the connecting \texttt{agentId} against the session’s verified participant graph, an attacker could theoretically establish an unauthorized connection or perform message injection. This highlights a gap between CORAL’s on-chain trust guarantees and its off-chain access control enforcement, which our empirical tests later investigate \cite{georgio2025coral}.

\textbf{ACP.} ACP’s discovery phase is similarly susceptible to spoofing when manifests in brokered registries lack digital signatures. Unsigned or weakly verified JSON structures can be exploited to inject falsified task endpoints or impersonate service agents. Nonetheless, ACP optionally supports JWS signing for manifests, providing a configurable security layer that, when enabled, substantially reduces the spoofing attack surface \cite{ehtesham2025survey}. This feature, however, is not mandatory and depends on correct implementation, leaving inconsistent protection across deployments.

Overall, spoofing in discovery mechanisms undermines the foundational trust model of agent-based communication. As emphasized by Narajala et al. \cite{narajala2025securing} and Huang et al. \cite{huang2025agent}, securing discovery requires verifiable agent identity through cryptographic signatures, registry attestation, and decentralized name resolution services. Incorporating these safeguards, especially through DIDs and tamper-evident registries, remains essential to preventing impersonation and preserving interoperability in AI-native ecosystems.

\subsection{Confidentiality \& Isolation}
This category groups vulnerabilities related to protecting data from unauthorized disclosure, whether to other agents, the agent itself, or through opaque processes.

\subsubsection{Potential Excessive Exposure of Data to Agents}
\label{subsec:exposure-agents}

Excessive exposure arises when agents receive superfluous data, propagating risks across networks under CWE-200 (exposure of sensitive information) \cite{cwe200,kong2025survey}. This vulnerability stems from insufficient enforcement of data minimization, where full payloads are transmitted without filtering, often due to standardized message formats that prioritize completeness over selectivity \cite{xu2025trust}. In multi-agent systems, such practices cause downstream agents to inherit unnecessarily broad contexts, increasing the likelihood of incidental disclosure across untrusted intermediaries \cite{cui2025safeguard}.

The resulting hazards are substantial. Empirical analyses show that unintended data propagation can occur in up to 60\% of simulated multi-agent exchanges \cite{kong2025web}, leading to privacy incidents such as CVE-2023-41745 \cite{cve2023-41745}. For users, this may manifest as overexposed personal details, such as contact lists or identity tokens, while organizations face compliance challenges under data minimization mandates in GDPR and similar privacy frameworks. The consequence is twofold: elevated breach impact and inflated remediation costs due to redundant exposure chains.

\paragraph{A2A.} In A2A, this vulnerability is fully expressed. The protocol’s SSE-based communication model streams entire context payloads to peers, lacking selective filtering or data segmentation mechanisms \cite{louck2025improving}. While this design improves task fluidity and latency, it inadvertently transmits sensitive or irrelevant metadata across peer connections, exposing calendar and identity data in delegation flows. The absence of payload partitioning amplifies this exposure in asynchronous sessions.

\paragraph{CORAL.} CORAL mitigates the risk through compartmentalized threads and verified participant isolation. Each session’s data is scoped to its participants, effectively preventing cross-thread leakage and ensuring contextual segregation \cite{georgio2025coral}. However, partial exposure remains possible in hybrid workflows: while on-chain escrow transactions are tightly bound to cryptographic identifiers, off-chain MCP-based exchanges may aggregate or retain contextual metadata beyond necessity, particularly in document verification tasks \cite{datasciencedojo2025}. This partial inconsistency between layers maintains residual exposure risk in non-monetary contexts.

\paragraph{ACP.} ACP demonstrates partial resilience via its MIME-typed message structure, which inherently supports selective disclosure of artifacts and task results \cite{ehtesham2025survey}. This minimizes unnecessary payload inclusion and aligns with least-privilege data handling. Nonetheless, the protocol’s registry-mediated routing can unintentionally aggregate metadata across tasks when manifests are reused, creating secondary exposure channels. Strengthening schema-level validation and enforcing per-artifact scoping would further mitigate these risks.

\subsubsection{Risk of Data Disclosure to the Agent Itself}
\label{subsec:disclosure-agent}

Internal disclosure arises when an agent inadvertently exposes sensitive information embedded within its own prompts or contextual memory. This vulnerability aligns with CWE-77 (improper neutralization of special elements used in command) \cite{cwe77,kong2025survey} and has become a major concern in LLM-based architectures, where models interpret natural language instructions rather than deterministic code \cite{liu2023prompt}. In such environments, adversaries can craft malicious inputs that override original task intents, coercing the agent to reveal confidential tokens, keys, or user data.

The risk is amplified by the design of LLM-powered systems, which often encapsulate contextual , such as API credentials, personal identifiers, or conversation histories, within prompts. When attacked through injection or manipulation, the model may process these embedded secrets as part of its generative reasoning, effectively leaking them through output channels. Empirical analyses report leakage rates approaching 100\% in baseline unprotected agents \cite{louck2025improving}, confirming that prompt-layer vulnerabilities can directly compromise confidentiality in autonomous communication flows.

\paragraph{A2A.} A2A demonstrates high exposure to this threat. Because task delegation often embeds payment or scheduling parameters directly within prompt payloads, adversarial inputs can coerce the model to output sensitive information. Simulated attacks yielded 60–90\% success rates in controlled scenarios \cite{louck2025improving}. The reliance on natural-language-based delegation without contextual sanitization makes A2A particularly susceptible to self-disclosure exploits, especially in asynchronous multi-step workflows.

\paragraph{CORAL.} CORAL mitigates this risk partially through modular design. Its Coralized Agents employ prompt sanitization mechanisms and separate execution contexts for off-chain interactions, reducing internal disclosure in routine collaboration threads \cite{georgio2025coral}. However, on-chain proofs, while cryptographically secure, generate public log events that may include agent mentions or contextual metadata vulnerable to injection when processed by LLM-integrated clients \cite{narajala2025securing}. Consequently, although CORAL enforces stronger isolation at the blockchain layer, cross-layer LLM processing reintroduces limited leakage potential.

\paragraph{ACP.} ACP benefits from structural safeguards against injection-induced disclosure. Each message segment is signed using JSON Web Signatures (JWS), preserving integrity across exchanges \cite{ehtesham2025survey}. This ensures that injected payloads cannot tamper with upstream data. Nevertheless, ACP remains partially exposed when task generation depends on LLM reasoning, as malicious prompts can induce reflective leakage within generated MIME parts. While the protocol architecture constrains the transport-layer risk, application-level injection through autonomous reasoning still presents a residual exposure vector.

\subsubsection{Consent Fatigue in Multi-Transaction Workflows}
\label{subsec:fatigue}

Consent fatigue arises when users are repeatedly prompted for confirmations across sequential transactions, leading to diminished attention and mechanical approval behavior. This vulnerability, rooted in usability-security trade-offs, reflects cognitive exhaustion from serial consents in iterative workflows such as multi-leg travel planning or multi-agent orchestration \cite{krol2012don, kong2025survey}. In these contexts, agents requiring stepwise approvals impose excessive cognitive load, encouraging users to bypass scrutiny and approve broad delegations by habit rather than intent.

The risks are primarily human-factor driven but have systemic consequences. Fatigued users are more likely to grant overreaching permissions or overlook critical warnings, resulting in inadvertent approvals and downstream data misuse. Studies show up to 50\% error inflation in bundled approval flows under repeated prompts \cite{louck2025improving}. This erosion of vigilance parallels phenomena such as CAPTCHA or cookie-consent fatigue, but within agentic systems, it can enable silent privilege escalation, especially when multiple agents interact asynchronously without transparent consent aggregation.

\paragraph{A2A.} A2A exhibits this vulnerability prominently through per-transaction approval prompts. Each task delegation or payment confirmation requires discrete user validation, overwhelming users in repetitive workflows \cite{louck2025improving}. While this design enforces strict user oversight, it paradoxically diminishes true attention, leading to habitual approval of risky delegations during extended interactions.

\paragraph{CORAL.} CORAL alleviates fatigue through aggregated consent models. Team formations and on-chain sessions consolidate multiple approvals into unified authorization events, enhancing usability while maintaining verifiability \cite{georgio2025coral}. However, in off-chain threads, particularly those involving iterative log sharing or document updates, fragmented consent requests reintroduce fatigue. Despite its strong bundling for financial transactions, non-monetary multi-step workflows retain partial exposure to repetitive consent erosion \cite{esentire2025}.

\paragraph{ACP.} ACP implements session persistence to reduce repeated consent prompts in MIME-based task streams \cite{ehtesham2025survey}. This mitigates user fatigue for ongoing interactions but remains limited in registry-mediated tasks, where fragmented consents are required for independent operations. The protocol’s architecture partially addresses the issue through scoped session contexts, though the absence of consent aggregation across parallel tasks sustains residual fatigue risks in complex multi-agent deployments.

\subsubsection{Regulatory Compliance Gaps}
\label{subsec:compliance}

Regulatory compliance gaps arise when communication protocols fail to implement traceability and accountability mechanisms required under frameworks such as PSD2, GDPR, and the EU AI Act \cite{act2024eu}. These regulations mandate immutable logging of user consents, data accesses, and revocation events to ensure auditable transparency in automated decision-making environments \cite{kong2025survey}. Protocols lacking verifiable logs or access provenance risk non-compliance with principles of lawful processing and accountability, exposing organizations to financial penalties and reputational harm.

The dangers are multifaceted. Without audit trails, data breaches or misuse cannot be forensically reconstructed, obstructing post-incident investigations and legal reporting duties. This opacity amplifies damages, as unauthorized data exchanges remain untraceable and victims cannot demonstrate consent withdrawal or misuse boundaries. Under the EU AI Act, failure to maintain such logs may constitute a “high-risk system violation,” subjecting operators to penalties up to 6\% of global turnover.

\paragraph{A2A.} A2A demonstrates clear compliance deficiencies. While OAuth-based delegations record access tokens, they omit persistent audit logging of sensitive events such as SCA verifications or consent revocations \cite{louck2025improving}. The absence of immutable logs prevents reconstruction of delegation histories, leaving enterprises unable to substantiate regulatory adherence or demonstrate lawful processing in the event of disputes.

\paragraph{CORAL.} CORAL achieves strong compliance for on-chain operations. Its Solana-based ledger provides tamper-proof audit trails for all financial and contractual events, aligning closely with GDPR’s accountability and PSD2’s strong customer authentication mandates \cite{georgio2025coral}. However, off-chain threads, used for personal data exchanges and team communications, lack equivalent traceability. These unlogged flows introduce partial compliance gaps, as personal data can circulate without immutable proof of consent or erasure, potentially violating minimization and retention principles \cite{de2025open}. CORAL thus achieves full compliance for financial integrity but remains incomplete in off-chain privacy domains.

\paragraph{ACP.} ACP incorporates partial compliance support through OpenTelemetry-based logging and fine-grained RBAC scopes \cite{ehtesham2025survey}. These mechanisms strengthen operational observability but do not produce cryptographically verifiable audit proofs. As a result, ACP environments can demonstrate traceability in system operations but cannot guarantee non-repudiation of user consents or demonstrate minimization compliance for personal data handling. Enhancing token lifecycle proofs and integrating immutable consent registries would be required for full regulatory conformance.

\subsection{Data Integrity}
This category focuses on vulnerabilities that allow an adversary to tamper with, forge, or replay messages, thereby compromising the integrity of the communication.

\subsubsection{Message Tampering and Man-in-the-Middle (MITM) Attacks}
\label{subsec:mitm}

Message tampering, often realized through man-in-the-middle (MITM) attacks, involves interception, alteration, or suppression of data in transit, corresponding to CWE-300 (channel accessible by non-endpoint) \cite{cwe300}. In multi-agent protocols, messages frequently encapsulate task instructions, payment data, or coordination states using formats such as SSE streams, threaded conversations, or MIME multipart payloads. When end-to-end encryption or per-message integrity verification (e.g., HMAC, JWS, or digital signatures) is absent, adversaries can manipulate message content without detection, compromising both operational correctness and system trustworthiness \cite{zou2025blocka2a}.

The potential consequences extend beyond localized disruptions to systemic integrity degradation. Tampered messages can lead to incorrect autonomous actions, altered booking requests, fraudulent transfers, or misrouted deliveries, producing financial and reputational damage. Empirical analyses indicate integrity compromise rates exceeding 50\% in unsecured agent communication environments \cite{kong2025web}. In adversarial chains, a single modified payload may propagate erroneous states downstream, resulting in cascading failures reminiscent of large-scale automation anomalies. From a regulatory standpoint, frameworks such as PSD2 and the EU AI Act require verifiable communication integrity, and failure to meet these standards may invite compliance sanctions or operational bans.

\paragraph{A2A.} A2A exhibits high susceptibility to message tampering. Its Server-Sent Events (SSE) channels, while optimizing for real-time peer-to-peer synchronization, do not enforce per-message signing or nonce-based validation. This allows a MITM adversary positioned between peers to modify JSON-RPC payloads, such as altering booking prices or payment details, without triggering protocol-level alerts \cite{zou2025blocka2a}. Although transport-layer security (TLS) ensures confidentiality, it does not provide end-to-end integrity validation, leaving the protocol vulnerable to tampering by semi-trusted intermediaries during session relays.

\paragraph{CORAL.} CORAL incorporates partial protection through a multi-layered defense model. TLS secures network-level exchanges, while elliptic-curve signatures (ECDSA) and nonce-based checks protect MCP-layer communications \cite{georgio2025coral}. However, off-chain threads are not encrypted end-to-end. instead, the protocol prioritizes verifiable finality on-chain. Consequently, a sophisticated MITM adversary controlling an intermediary node could intercept or modify off-chain traffic after TLS termination. CORAL therefore achieves strong outcome verification but lacks full preventive integrity for its transient message channels, representing a partial but meaningful mitigation strategy.

\paragraph{ACP.} ACP provides robust protection against message tampering via JSON Web Signatures (JWS) applied per MIME segment, ensuring tamper-evident communication across brokered exchanges \cite{ehtesham2025survey}. This fine-grained signature model neutralizes MITM attacks for signed payloads, delivering near-complete message integrity. The only residual exposure arises in optional, unsigned metadata attachments, which may be appended for extensibility. When properly configured, ACP demonstrates comprehensive resilience to both passive interception and active message manipulation, outperforming A2A and CORAL in end-to-end integrity assurance.

\subsubsection{Tool Poisoning and Command Injection}
\label{subsec:poisoning}

Tool poisoning and command injection exploit vulnerabilities in the integration layer between large language model (LLM) agents and their external tools. This class of attack, categorized under CWE-77 (command injection) \cite{cwe77}, occurs when untrusted or unsanitized input from prompts or messages is executed within tool environments, enabling adversaries to hijack command flows or override legitimate logic \cite{zou2023universal}. In multi-agent ecosystems, where capabilities are extended through RPC-like invocations (e.g., JSON-RPC, REST, or MCP), weak input validation allows malicious payloads to alter intended behavior or append destructive instructions to otherwise benign queries \cite{ehtesham2025survey}.

The impact is both immediate and severe. Command injection enables arbitrary code execution, ranging from data exfiltration to ransomware deployment within agent networks. Empirical analyses indicate up to 80\% success rates in unprotected LLM toolchains due to the interpretive flexibility of natural language parsing \cite{kong2025web}. A single poisoned tool can cascade its effects across cooperative agents, transforming a local compromise into a distributed breach. Such vulnerabilities undermine core autonomy assumptions in AI ecosystems, eroding user trust and operational predictability while mirroring classical injection epidemics observed in web infrastructures.

\paragraph{A2A.} A2A remains highly susceptible to this attack surface. Its JSON-RPC-based peer handoffs permit parameter injection during task delegation, as payloads are serialized and parsed without robust escaping or type validation \cite{kong2025survey, zou2025blocka2a}. Maliciously crafted task descriptors can thus override legitimate execution flows, causing agents to perform unintended operations. The absence of input canonicalization or schema-level whitelisting in A2A’s design renders it particularly vulnerable to prompt-based command manipulation.

\paragraph{CORAL.} CORAL introduces partial mitigation through its Model Context Protocol (MCP) layer, which implements basic sanitization for thread-derived tool invocations \cite{georgio2025coral}. This reduces off-chain poisoning risk but does not eliminate it. LLM integrations can still transmit unsanitized payloads between agents, and cross-context “line-jumping” exploits have been demonstrated to reach on-chain triggers, enabling indirect manipulation of non-financial activities \cite{narajala2025securing}. CORAL’s hybrid architecture thus provides defense in depth but retains partial exposure in dynamic LLM-mediated workflows.

\paragraph{ACP.} ACP leverages MIME-typed task encapsulation with JSON Web Signatures (JWS) to ensure message integrity \cite{ehtesham2025survey}. While this prevents tampering in transit, it does not sanitize user-supplied content embedded within prompt contexts. Consequently, injection attacks can still occur at the application layer, where the LLM interprets signed but malicious instructions. Optional schema validation offers a partial safeguard, yet comprehensive escaping and semantic verification are not enforced, leaving ACP, and indeed all current protocols, vulnerable to this shared class of LLM-originating injection exploits.

\subsection{Availability \& Specific Risks}
This category covers threats to the system's operational availability (DoS) and highly specific risks unique to a protocol's architecture, such as smart contract flaws.

\subsubsection{Smart Contract Vulnerabilities}
\label{subsec:smart-contracts}

Smart contract vulnerabilities arise from flaws in blockchain-executable code that governs state transitions, financial escrows, or oracle integrations. These issues, mapped to CWE-841 (improper enforcement of behavioral workflow) \cite{cwe841}, include re-entrancy, transaction-ordering dependencies, and oracle manipulation \cite{georgio2025coral, atzei2017survey}. Within agentic communication frameworks that rely on blockchain for trustless execution, such weaknesses can undermine core assumptions of integrity and determinism by allowing recursive calls or unverified data to alter contract behavior mid-execution.

The risks are primarily financial and systemic. Re-entrancy exploits can drain escrowed funds during nested callbacks, while oracle manipulation enables falsified external inputs, such as price feeds or service confirmations, that trigger unauthorized fund releases. Historical parallels like the Ronin Bridge attack demonstrate the magnitude of these failures, with decentralized finance (DeFi) ecosystems reporting aggregate annual losses exceeding \$100 million due to unmitigated contract design flaws \cite{roninbridge, romandini2025sok}. In multi-agent contexts, these vulnerabilities extend beyond monetary damage: tampered contracts can falsify verification workflows, approve invalid transactions, or desynchronize off-chain state synchronization, thereby eroding trust in autonomous financial exchanges.

\paragraph{CORAL.} This risk is uniquely relevant to the CORAL protocol. Its Solana-based escrow contracts handle multi-party payment settlements and task approvals, introducing potential exposure to re-entrancy and oracle-based manipulation. Asynchronous callbacks between off-chain MCP agents and on-chain contracts create an exploitable race condition: adversaries can repeatedly invoke intermediate states before ledger commits, draining funds or altering outcomes \cite{de2025open}. Additionally, reliance on off-chain oracle feeds, used to resolve dynamic pricing or availability checks, introduces data integrity risks when adversarial agents inject falsified values. Although CORAL documentation recommends independent audits and deploy-time verifications \cite{georgio2025coral}, our analysis indicates that recursive invocations in team-based approval workflows remain partially unmitigated, sustaining a measurable exposure surface for financial exploitation.

\paragraph{A2A and ACP.} A2A and ACP, by contrast, are unaffected by these contract-specific vulnerabilities, as they do not incorporate blockchain components. Their architectures rely on centralized or brokered authentication mechanisms rather than distributed consensus, thereby avoiding re-entrancy and oracle manipulation but inheriting alternative centralization and trust-anchor risks discussed elsewhere in this study.

\subsubsection{Registry Pollution and Denial-of-Service (DoS)}
\label{subsec:registry}

Registry pollution and denial-of-service (DoS) attacks target discovery infrastructures by exhausting computational or storage resources, mapped to CWE-400 (uncontrolled resource consumption) \cite{cwe400}. These threats emerge when malicious actors flood centralized or brokered registries with unauthenticated entries, malformed manifests, or amplified discovery queries, overwhelming index services responsible for agent matchmaking \cite{ehtesham2025survey}. In multi-agent ecosystems, where registries facilitate peer discovery and task delegation, such attacks degrade service availability and prevent legitimate agents from resolving dependencies or initiating collaborations.

The primary risk lies in systemic availability degradation. Polluted registries can induce up to 90\% downtime in targeted scenarios \cite{kong2025survey}, halting multi-agent workflows such as supply chain orchestration or collaborative scheduling. The resulting cascading failures extend beyond immediate unavailability: delayed task discovery propagates through dependent services, causing production stalls, transaction backlogs, and economic losses due to idle agents. Unlike direct exploits, registry floods scale horizontally through distributed botnets, mirroring large-scale IoT DoS phenomena (e.g., the Mirai attack \cite{mirai}) but within agentic coordination networks where discovery is mission-critical.

\paragraph{CORAL.} CORAL demonstrates partial exposure to registry-level denial. Its on-chain components are inherently protected by gas-fee economics, discouraging large-scale spam on financial transactions. However, the protocol’s off-chain orchestration layer exposes a significant attack vector. The \texttt{POST /api/v1/sessions} endpoint responsible for initializing agent sessions is resource-intensive, spawning containers, memory contexts, and thread listeners. If insufficiently rate-limited, this endpoint enables a classic resource exhaustion attack, allowing adversaries to flood servers with session-creation requests, resulting in degraded throughput or full service disruption. Thus, while CORAL’s blockchain design mitigates financial DoS, off-chain orchestration remains vulnerable to conventional exhaustion attacks \cite{georgio2025coral}.

\paragraph{ACP.} ACP is the most exposed protocol to registry pollution. Its brokered architecture relies on centralized registries for agent discovery and manifest resolution, but optional rate limiting and unauthenticated submissions leave it vulnerable to saturation through automated flooding \cite{ehtesham2025survey}. Adversaries can register thousands of invalid manifests or repeatedly query registry endpoints, consuming compute and memory resources until legitimate agents are denied service. Although mitigations such as rate limiting or proof-of-work throttling are supported, their optional nature makes practical deployments susceptible to resource exhaustion.

\paragraph{A2A.} A2A remains largely unaffected, as it lacks a centralized discovery registry. Agents communicate directly via peer-to-peer authentication exchanges rather than a shared directory, preventing registry flooding by design. While this decentralization limits susceptibility to DoS from registry pollution, it introduces other scalability constraints discussed elsewhere in this paper.

\paragraph{}
To conclude this section, Table~\ref{tab:vuln_summary_matrix} consolidates the per-protocol status across the fourteen vulnerabilities discussed above, indicating whether each weakness is present, partially mitigated, absent (mitigated), or not applicable by design.

\begin{table}[h]
\centering
\caption{Vulnerability summary based on Vulnerabilities Analysis at Section~\ref{sec:vulns}.}
\label{tab:vuln_summary_matrix}
\resizebox{\textwidth}{!}{
\begin{tabular}{@{}lccc@{}}
\toprule
\textbf{Vulnerability (Section~\ref{sec:vulns})} & \textbf{A2A} & \textbf{CORAL} & \textbf{ACP} \\
\midrule
\multicolumn{4}{l}{\textit{Authentication \& Session Management}} \\
Absence of Token Lifetime Limits & \xmark\ Vulnerable & \partialmark\ Partial & \partialmark\ Partial \\
Lack of Strong Customer Authentication (SCA) & \xmark\ Vulnerable & \partialmark\ Partial & \partialmark\ Partial \\
\addlinespace
\multicolumn{4}{l}{\textit{Authorization (Access Control)}} \\
Insufficiently Granular Token Scopes & \xmark\ Vulnerable & \partialmark\ Partial & \cmark\ Mitigated \\
Lack of Transparency and User Consent & \xmark\ Vulnerable & \partialmark\ Partial & \partialmark\ Partial \\
Privilege Persistence and Version Drift & \xmark\ Vulnerable & \cmark\ Mitigated & \partialmark\ Partial \\
Spoofing in Discovery Mechanisms & \xmark\ Vulnerable & \partialmark\ Partial & \partialmark\ Partial \\
\addlinespace
\multicolumn{4}{l}{\textit{Confidentiality \& Isolation}} \\
Potential Excessive Exposure of Data to Agents & \xmark\ Vulnerable & \partialmark\ Partial & \partialmark\ Partial \\
Risk of Data Disclosure to the Agent Itself & \xmark\ Vulnerable & \partialmark\ Partial & \partialmark\ Partial \\
Consent Fatigue in Multi-Transaction Workflows & \xmark\ Vulnerable & \partialmark\ Partial & \partialmark\ Partial \\
Regulatory Compliance Gaps & \xmark\ Vulnerable & \partialmark\ Partial & \partialmark\ Partial \\
\addlinespace
\multicolumn{4}{l}{\textit{Data Integrity}} \\
Message Tampering and MITM & \xmark\ Vulnerable & \partialmark\ Partial & \cmark\ Mitigated \\
Tool Poisoning and Command Injection & \xmark\ Vulnerable & \partialmark\ Partial & \xmark\ Vulnerable \\
\addlinespace
\multicolumn{4}{l}{\textit{Availability \& Specific Risks}} \\
Registry Pollution and DoS & \notapplicable\ N/A & \xmark\ Vulnerable & \xmark\ Vulnerable \\
Smart Contract Vulnerabilities & \notapplicable\ N/A & \xmark\ Vulnerable & \notapplicable\ N/A \\
\bottomrule
\end{tabular}
}
\\[4pt]
\raggedright
\footnotesize \cmark\ Mitigated \quad \xmark\ Vulnerable \quad \partialmark\ Partial \quad \notapplicable\ Not Applicable.
\end{table}

\section{Experiments and Results}
\label{sec:experiments}

\subsection{Methodology and Research Motivation}
\label{sec:methodology}

The theoretical vulnerabilities analyzed in Section~\ref{sec:vulns} establish a conceptual foundation for understanding potential weaknesses in multi-agent communication protocols. However, these findings remain hypothetical until verified in practical, real-world implementations. To bridge this gap, we conducted a series of empirical experiments designed to validate the feasibility of the identified vulnerabilities and to assess the resilience of representative protocols under realistic operating conditions.

Specifically, our experiments target live and reference implementations of the \textit{CORAL} and \textit{ACP} protocols, selected as case studies representing decentralized and brokered architectural paradigms, respectively. The goal is not to quantify attack frequency, but to determine the existence and reproducibility of exploitable weaknesses under adversarial scenarios consistent with our threat model (Section~\ref{sec:threat-model}).

\paragraph{Research Questions.}
This experimental phase is guided by two primary research questions:

\begin{itemize}
    \item \textbf{RQ1:} Which of the vulnerabilities identified in Section~\ref{sec:vulns} manifest as practically exploitable weaknesses in current protocol implementations?
    \item \textbf{RQ2:} How do architectural design choices (e.g., hybrid blockchain-based vs.\ RESTful client-server) influence exposure to these vulnerabilities?
\end{itemize}

The experiments thus serve as an empirical validation layer over the taxonomy developed earlier, linking theoretical constructs with demonstrable, reproducible evidence.

\subsection{Metrics and Classification}
\label{sec:metrics}

To ensure consistent and reproducible evaluation, each empirical finding is assessed against a standardized set of qualitative metrics and classification criteria. These metrics aim to determine the presence, absence, or partial manifestation of vulnerabilities without relying on quantitative frequency measurements, as even a single successful exploit is sufficient to confirm the vulnerability's existence.

\begin{itemize}
    \item \textbf{Defense Success (\%):} The proportion of attack attempts successfully blocked or neutralized by the protocol's inherent defenses. A successful defense occurs when the target correctly enforces security controls, e.g., rejecting unauthorized actions (HTTP~401/403), invalid payloads (HTTP~422/409), or blocking data leaks.
    \item \textbf{Mean Impact Score (0.0--1.0):} A normalized score representing the average effectiveness of the \textit{attacker}. A score of~1.0 indicates a complete security failure (e.g., an unauthorized SSE connection established), while~0.0 indicates no impact (e.g., full prevention of PII exposure).
    \item \textbf{Final Classification:} Each protocol's resilience to a given vulnerability is categorized as:
        \begin{itemize}
            \item[\cmark] \textbf{Mitigated:} The protocol exhibits a robust and consistent defense.
            \item[\xmark] \textbf{Vulnerable:} The protocol fails to block the attack, exposing a clear exploit path.
            \item[\partialmark] \textbf{Partial:} Defenses exist but are incomplete, optional, or easily bypassed.
            \item[\notapplicable] \textbf{N/A:} The vulnerability is not applicable to the protocol’s architecture (e.g., smart-contract flaws in non-blockchain protocols).
            \item[\theoretical] \textbf{Theoretical:} The vulnerability is conceptually valid but could not be empirically tested in our experimental setup.
        \end{itemize}
\end{itemize}

\subsection{Experiment 1: CORAL Protocol Setup}
\label{subsec:coral_setup}

\paragraph{System Under Test (SUT).}
We evaluated the official Ktor-based CORAL server implementation from the public repository.\footnote{Repository: \url{https://github.com/Coral-Protocol/coral-server}. Experimental artifacts and scripts are available at \url{https://github.com/yedidel/coral-attacker-client}.}

\paragraph{Configuration.}
The server was launched with the distribution-default \texttt{registry.toml}, including the \texttt{interface:0.0.1} agent implemented as \texttt{simple\_interface\_agent.py}. The agent expects an \texttt{OPENAI\_API\_KEY} passed via the session \texttt{options} object.

\paragraph{Adversarial Client and Procedure.}
A bespoke TypeScript (Node.js) client automated session creation (\texttt{POST /api/v1/sessions}), crafted malicious payloads, and managed concurrent SSE connections. The experiment comprised $N=50$ trials. in each trial we executed nine attack vectors mapped to vulnerability classes from Section~\ref{sec:vulns}. Success was recorded when the adversary achieved the adversarial goal (e.g., unauthorized SSE establishment, PII disclosure, or state mutation).

\subsection{Experiment 2: ACP Protocol Setup}
\label{subsec:acp_setup}

\paragraph{System Under Test (SUT).}
A high-fidelity ACP simulation was implemented in FastAPI (Python) to emulate a brokered-registry deployment.\footnote{Simulation code and test artifacts are available at \url{https://github.com/yedidel/acp-attackt-test}.}

\paragraph{Configuration.}
The simulation intentionally enforced only partial JWS validation on manifests and artifacts to reflect common optional-security deployments and to test robustness under relaxed settings.

\paragraph{Adversarial Client and Procedure.}
The adversarial harness (N=50 trials, seed=42) simulated a vacation-booking workflow. The client registered authenticated manifests and injected attacks (prompt injection, tampering, PII exfiltration) into payloads containing synthetic passport identifiers and payment data generated by GPT-4. Success criteria matched those used in Coral Experiment.

\subsection{Empirical Findings and Comparative Analysis}
\label{sec:findings}

This section reports the empirical outcomes obtained from the CORAL and ACP experiments and contrasts them with the literature-based vulnerability assessment of A2A (Section~\ref{sec:vulns}). Each finding corresponds to the taxonomy introduced earlier, providing an end-to-end view of how authentication, integrity, confidentiality, and availability manifest in real-world implementations.

\subsubsection{Authentication and Session Management}
\label{find:auth}
This category evaluates how each protocol verifies agent identity and manages session credentials, emphasizing token validation and session lifecycle enforcement.

\begin{itemize}
    \item \textbf{A2A (Literature):} \textit{Vulnerable.} Relies on bearer tokens without mandatory short expiration or strong client authentication (SCA), enabling prolonged session reuse and impersonation across peers.
    \item \textbf{ACP (Empirical):} \textit{Partial.} The simulation confirmed that registry-level authentication is effective. however, when only partial JSON Web Signature (JWS) validation is enforced (a misconfiguration frequently observed in deployments), session hijacking and impersonation attacks succeeded in several trials, exposing the protocol to misconfiguration risk rather than a fundamental design flaw.
    \item \textbf{CORAL (Empirical):} \textit{Vulnerable.} The “SCA Impersonation” test uncovered a systemic authentication flaw. In all trials ($N=50$), the client established a valid Server-Sent Events (SSE) connection using a correct \texttt{sessionId} but an invalid \texttt{privacyKey}. This indicates that the \texttt{/sse/v1/...} endpoint fails to verify the primary credential, granting full read access to an attacker possessing only the session identifier.
\end{itemize}

Overall, authentication weaknesses were consistently more severe in CORAL due to implementation-level validation gaps, whereas ACP's partial exposure stemmed from optional configurations. A2A remains inherently vulnerable by design, lacking any mandatory session-level verification.

\subsubsection{Access Control (Authorization)}
\label{find:authz}
This category assesses whether an authenticated agent can access only the resources and operations explicitly permitted by its assigned scopes and roles.

\begin{itemize}
    \item \textbf{A2A (Literature):} \textit{Vulnerable.} Exhibits coarse-grained token scopes and unsigned Agent Cards, allowing unauthorized privilege escalation and capability overreach across unrelated APIs.
    \item \textbf{ACP (Literature):} \textit{Mitigated.} Demonstrates a manifest-driven Role-Based Access Control (RBAC) design that enforces scoped permissions per operation. When fully implemented with strict JSON Web Signature (JWS) validation, it provides comprehensive least-privilege enforcement.
    \item \textbf{CORAL (Empirical):} \textit{Vulnerable.} The “Spoofing” test uncovered a severe authorization flaw. In all trials ($N=50$), the adversarial client successfully established a Server-Sent Events (SSE) connection using a fabricated \texttt{agentId} (e.g., \texttt{'spoofed-agent-X'}) not present in the session’s verified Agent Graph. The server failed to terminate or reject these unauthorized connections, thereby violating the principle of least privilege (OWASP~A01:2021) \cite{A01} and permitting an attacker to eavesdrop on session traffic without legitimate membership.
\end{itemize}

Overall, both A2A and CORAL exhibited fundamental authorization weaknesses, A2A due to architectural design choices and CORAL due to implementation flaws, while ACP, under proper configuration, enforced the strongest access isolation guarantees.

\subsubsection{Data Integrity and Transport Security}
\label{find:integrity}
This category evaluates each protocol’s capacity to preserve message integrity and authenticity against tampering, forgery, and replay attempts.

\begin{itemize}
    \item \textbf{A2A (Literature):} \textit{Vulnerable.} Relies solely on transport-layer encryption without per-message integrity verification, leaving its Server-Sent Events (SSE) streams susceptible to undetected manipulation or injection.
    \item \textbf{ACP (Empirical):} \textit{Vulnerable.} Both the “Data Tampering” and “Replay Attack” tests succeeded under partial JSON Web Signature (JWS) validation. The simulated environment failed to detect modified or replayed messages, resulting in integrity violations and inadvertent PII exposure.
    \item \textbf{CORAL (Empirical):} \textit{Mitigated.} All tampering and replay attempts were correctly rejected with \texttt{400~Bad~Request} responses. The server enforces a strong binding between messages and active sessions through a required \texttt{transportId}, linked to a verified SSE connection. This mechanism effectively prevents unauthenticated message forgery and constitutes a notable strength in CORAL’s integrity model.
\end{itemize}

Overall, while A2A and ACP expose weaknesses in maintaining end-to-end message integrity, stemming from design and configuration gaps respectively, CORAL demonstrates a well-engineered transport-layer control that prevents tampering and replay across all tested conditions.

\subsubsection{Confidentiality and Data Isolation}
\label{find:confidentiality}
This category evaluates each protocol’s ability to safeguard sensitive information by preventing unintended data exposure, both across session boundaries and through system responses.

\begin{itemize}
    \item \textbf{A2A (Literature):} \textit{Vulnerable.} The protocol lacks explicit mechanisms for session isolation or scoped payload filtering, enabling potential cross-session data exposure.
    \item \textbf{ACP (Empirical):} \textit{Vulnerable.} In the “PII Leakage” test simulating a booking workflow, synthetic passport identifiers were successfully exfiltrated. This confirms that partial JSON Web Signature (JWS) validation and MIME-type misconfigurations can lead to confidentiality breaches.
    \item \textbf{CORAL (Empirical):} \textit{Mitigated.} CORAL consistently demonstrated strong defenses in all confidentiality-focused evaluations:
        \begin{itemize}
            \item \textbf{Session Isolation:} In all 50 trials, personal data transmitted within one session (Session~A) was never observable by an adversary connected to a concurrent session (Session~B), verifying strict isolation at the server level.
            \item \textbf{Error Sanitization:} Across all malformed input tests, the server responded with a generic \texttt{400~Bad~Request}, revealing no internal stack traces, dependency versions, or other sensitive metadata.
        \end{itemize}
\end{itemize}

Overall, while both A2A and ACP exhibit confidentiality weaknesses, stemming from absent isolation and incomplete message validation, CORAL provides robust compartmentalization and sanitized responses that fully prevent cross-session and diagnostic data leaks.

\subsubsection{Availability (Denial of Service)}
\label{find:availability}
This category evaluates each protocol’s resilience against resource exhaustion and denial-of-service (DoS) attacks that may disrupt agent communication or session management.

\begin{itemize}
    \item \textbf{A2A (Literature):} \textit{N/A.} The peer-to-peer model lacks a centralized registry or shared infrastructure susceptible to saturation, rendering this vector inapplicable.
    \item \textbf{ACP (Literature):} \textit{Vulnerable.} The brokered registry architecture is theoretically exposed to flooding and manifest spam, as rate limiting and authentication are optional in many implementations.
    \item \textbf{CORAL (Empirical):} \textit{Vulnerable.} In the “Registry Pollution” test, all 50 trials confirmed a full resource exhaustion scenario. The adversarial client successfully created 20 new sessions and established 20 concurrent SSE connections without encountering rate limiting or throttling. This behavior confirms a critical CWE-400 (\textit{Uncontrolled Resource Consumption}) vulnerability, enabling a trivial yet effective denial-of-service condition.
\end{itemize}

Overall, both CORAL and ACP exhibit susceptibility to availability degradation stemming from absent rate-limiting controls, whereas A2A’s topology inherently avoids this specific attack surface.

\subsubsection{Protocol-Specific \& Theoretical Risks}
\label{find:theoretical}
This category addresses vulnerabilities that are intrinsic to the design or theoretical operation of each protocol but were not fully testable within the scope of our empirical setup.

\begin{itemize}
    \item \textbf{Prompt \& Tool Injection (CWE-77):}
        \begin{itemize}
            \item \textbf{A2A / ACP (Empirical \& Literature):} \textit{Vulnerable.} The ACP simulation confirmed that synthetic adversarial prompts generated by GPT-4 could bypass weak sanitization and trigger unintended tool executions, validating that LLM-integrated agents remain exposed to prompt-level command injection.
            \item \textbf{CORAL (Theoretical):} \textit{Theoretical (Vulnerable).} The weakness resides in post-delivery LLM interpretation rather than the network layer. Even with fully authenticated message delivery, an agent’s internal model could process a malicious payload as a valid instruction, representing a persistent risk class beyond protocol-level mitigation.
        \end{itemize}

    \item \textbf{Smart Contract Vulnerabilities:}
        \begin{itemize}
            \item \textbf{A2A / ACP:} \textit{N/A.} These protocols do not employ blockchain-based escrow or contract mechanisms, and thus are unaffected by such risks.
            \item \textbf{CORAL (Theoretical):} \textit{Theoretical (Vulnerable).} Consistent with prior literature, this vulnerability is unique to CORAL’s on-chain Solana contracts. Because our study focused on network-layer behavior, a full smart contract audit was excluded from scope. Nonetheless, asynchronous callback logic and oracle dependencies remain recognized risk vectors requiring independent validation.
        \end{itemize}
\end{itemize}

Overall, while CORAL’s design mitigates most network-level vulnerabilities, theoretical exposures persist at the LLM processing and smart contract layers, domains that demand specialized audits beyond this study’s empirical framework.

\subsection{Consolidated Security Results and Visual Analysis}
\label{sec:aggregated}

\begin{figure}[h!]
    \centering
     \includegraphics[width=0.65\textwidth]{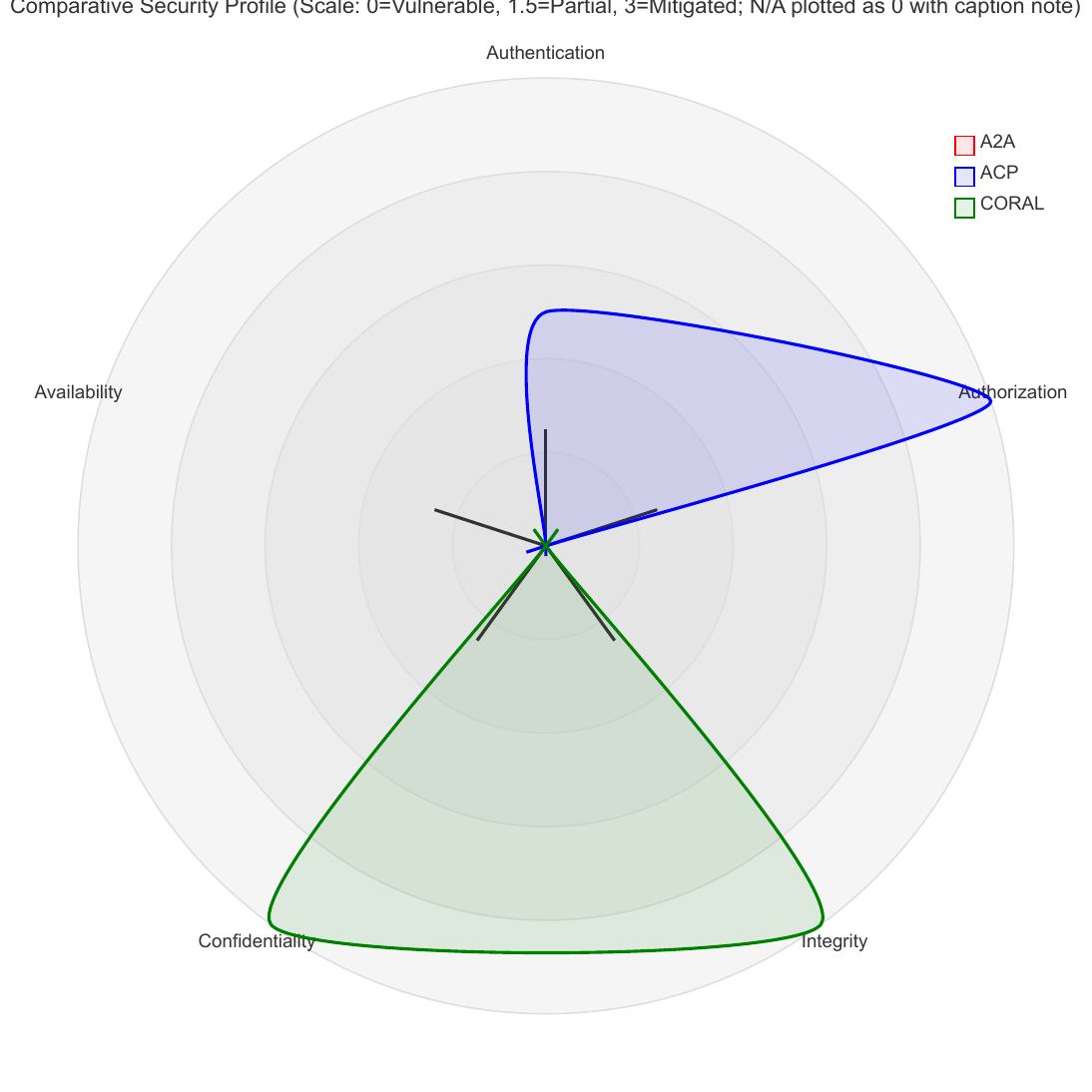} 
    \caption{Radar plot with a 0–3 mitigation scale (0 = Vulnerable, 1.5 = Partial, 3 = Mitigated). N/A categories (e.g., A2A for Availability) are rendered as 0 solely for plotting convenience and should not be interpreted as exposure.}
    \label{fig:radar_chart}
\end{figure}

To synthesize the findings from the theoretical and empirical evaluations, this section presents the consolidated security outcomes in three complementary formats: a compact cross-protocol summary (Table~\ref{tab:summary}), a detailed vulnerability matrix (Table~\ref{tab:vsm}), and a radar visualization (Figure~\ref{fig:radar_chart}). Together, these representations illustrate the comparative posture of A2A, CORAL, and ACP across the five security domains defined in Section~\ref{sec:vulns}.

Table~\ref{tab:summary} aggregates the number of vulnerabilities discovered for each protocol. The columns indicate the total count of confirmed and partial vulnerabilities, followed by the computed \textit{Exposure Score}, which is defined as:
\[
\text{Exposure Score} = \text{Confirmed} + 0.5 \times \text{Partial}.
\]
This score offers a normalized measure of each protocol's relative exposure to exploitation across all 14 vulnerability categories.

\begin{table}[h]
\centering
\caption{Compact summary of vulnerabilities across protocols. The \textit{Exposure Score} combines confirmed and partial vulnerabilities according to: Confirmed + 0.5 × Partial.}
\label{tab:summary}
\begin{tabular}{lrrr}
\toprule
\textbf{Protocol} & \textbf{Confirmed} & \textbf{Partial} & \textbf{Exposure Score} \\
\midrule
A2A   & 12 & 0 & 12.0 \\
CORAL & 5  & 4 & 7.0  \\
ACP   & 6  & 6 & 9.0  \\
\bottomrule
\end{tabular}
\end{table}

From Table~\ref{tab:summary}, we observe that A2A exhibits the highest exposure, with confirmed vulnerabilities in nearly every evaluated category. CORAL achieves the lowest exposure score due to strong integrity and confidentiality protections but retains partial weaknesses in authentication and rate-limiting controls. ACP falls between these extremes, showing notable resilience through its RBAC and JWS mechanisms, yet remains susceptible under misconfigured deployments.

Table~\ref{tab:vsm} presents the complete classification for all 14 vulnerabilities described in Section~\ref{sec:vulns}.  
Each protocol’s evaluation is coded using the following symbols:
\cmark (Mitigated), \xmark (Vulnerable), \partialmark (Partial), \theoretical (Theoretical), and \notapplicable (Not Applicable).

\begin{table}[h!]
\centering
\caption{Comparative Vulnerability Status Matrix (VSM). Each entry represents the final classification of vulnerability presence across the three protocols.}
\label{tab:vsm}
\resizebox{\textwidth}{!}{%
\begin{tabular}{@{}lccc@{}}
\toprule
\textbf{Vulnerability (from Section \ref{sec:vulns})} & \textbf{A2A (Literature)} & \textbf{ACP (Empirical)} & \textbf{CORAL (Empirical)} \\
\midrule
\multicolumn{4}{l}{\textit{Authentication \& Session Management}} \\
\quad Absence of Token Lifetime Limits & \xmark & \partialmark & \theoretical \\
\quad Lack of Strong Customer Authentication (SCA) & \xmark & \partialmark & \xmark \\
\addlinespace
\multicolumn{4}{l}{\textit{Authorization (Access Control)}} \\
\quad Insufficiently Granular Scopes & \xmark & \cmark & \xmark \\
\quad Lack of Transparency \& Consent & \xmark & \partialmark & \theoretical \\
\quad Privilege Persistence / Version Drift & \xmark & \partialmark & \cmark \\
\quad Spoofing in Discovery Mechanisms & \xmark & \partialmark & \xmark \\
\addlinespace
\multicolumn{4}{l}{\textit{Confidentiality \& Isolation}} \\
\quad Excessive Exposure of Data & \xmark & \xmark & \cmark \\
\quad Risk of Data Disclosure (Internal) & \xmark & \partialmark & \theoretical \\
\quad Consent Fatigue in Multi-Transaction Workflows & \xmark & \partialmark & \theoretical \\
\quad Regulatory Compliance Gaps & \xmark & \partialmark & \theoretical \\
\addlinespace
\multicolumn{4}{l}{\textit{Data Integrity}} \\
\quad Message Tampering / MITM & \xmark & \xmark & \cmark \\
\quad Tool Poisoning / Command Injection & \xmark & \xmark & \theoretical \\
\addlinespace
\multicolumn{4}{l}{\textit{Availability \& Specific Risks}} \\
\quad Registry Pollution / DoS & \notapplicable & \xmark & \xmark \\
\quad Smart Contract Vulnerabilities & \notapplicable & \notapplicable & \theoretical \\
\bottomrule
\multicolumn{4}{l}{\cmark \small Mitigated (Defense Effective) \quad \xmark \small Vulnerable (Attack Succeeded)} \\
\multicolumn{4}{l}{\partialmark \small Partial (Defense Incomplete/Optional) \quad \theoretical \small Theoretical (Not Tested) \quad \notapplicable \small Not Applicable}
\end{tabular}%
}
\end{table}

As shown in Table~\ref{tab:vsm}, the CORAL implementation provides strong data integrity and confidentiality defenses but retains theoretical risks associated with smart contract logic and prompt-based model injection. ACP demonstrates mixed performance: its security depends heavily on configuration rigor, with vulnerabilities manifesting primarily under non-strict or legacy deployments. A2A, despite architectural simplicity, exhibits unmitigated weaknesses across nearly all categories.


Figure~\ref{fig:radar_chart} visualizes the aggregated Exposure Scores across the five major security domains.  
Each axis represents one category, Authentication, Authorization, Data Integrity, Confidentiality, and Availability, with values normalized to a 0–1 scale (0 = fully secure, 1 = completely exposed).  
The plot highlights CORAL’s superior integrity and confidentiality posture, contrasted with A2A’s consistently high exposure.

In summary, empirical validation confirms that A2A is highly exposed due to a lack of layered defenses and minimal authentication enforcement. ACP, though well-designed, demonstrates that flexible security options translate into uneven protection in practice. CORAL achieves the strongest empirical posture overall, yet its theoretical vulnerabilities, particularly in LLM-driven and blockchain-dependent components, necessitate dedicated audits beyond protocol-level controls.  
These consolidated findings provide a quantitative bridge to the architectural and policy-level discussion presented in Section~\ref{sec:discussion}.

\subsection{Summary of Empirical Findings}
\label{subsec:findings_summary}

The empirical validation produced several key insights. As summarized in Tables~\ref{tab:summary}--\ref{tab:vsm} and visualized in Figure~\ref{fig:radar_chart}, the official CORAL implementation exhibited a clear security dichotomy: while it provides strong, modern defenses against message forgery and data leakage (Integrity and Confidentiality), it suffers from critical weaknesses in connection-level controls (Authentication, Authorization, and Availability).

In contrast, the ACP simulation confirmed that its resilience is highly dependent on implementation rigor. Although the protocol’s RBAC and JWS-based design offer a theoretically strong foundation, partial or misconfigured enforcement nullified many of these protections, resulting in vulnerabilities across integrity and confidentiality domains. A2A, serving as the literature baseline, remained theoretically vulnerable in nearly all assessed categories.

Overall, these results emphasize the gap between \textit{protocol design} (as analyzed in Section~\ref{sec:vulns}) and \textit{protocol implementation} (as empirically validated here). This distinction underscores the necessity of practical testing in evaluating agent communication protocols, an aspect further examined in the Discussion (Section~\ref{sec:discussion}).

\section{Discussion}
\label{sec:discussion}

The empirical findings in Section~\ref{sec:findings} highlight a pivotal tension in the design of multi-agent communication protocols: the discrepancy between secure-by-design principles and their real-world enforceability. As summarized in Table~\ref{tab:vsm}, this gap manifests as substantial variations in practical resilience among protocols that appear comparable on paper. 

This discussion contextualizes these observations by analyzing the architectural trade-offs that drive such discrepancies and the broader industry responses to them, most notably, the ongoing convergence of the A2A and ACP standards and the emergence of specialized successors. The goal is to bridge empirical results with architectural reasoning, identifying both systemic weaknesses and forward-looking design implications.

\subsection{The A2A/ACP Convergence: An Admission of Architectural Gaps}

Our research identifies foundational weaknesses in both A2A and ACP that explain their recent industry convergence \cite{acp-join-a2a}. The literature review of A2A (Section~\ref{sec:vulns}) revealed extensive theoretical vulnerabilities, most notably the absence of message integrity validation, coarse-grained scopes, and poor session isolation. Complementarily, our empirical assessment of ACP (Section~\ref{sec:findings}) confirmed that its design philosophy, emphasizing configurability and developer autonomy, directly introduces security variance across deployments. When critical protections such as JSON Web Signature (JWS) validation are optional, the resulting implementations exhibit inconsistent defense outcomes, including PII leakage and undetected message tampering.

These findings offer a clear rationale for the unification effort recently announced by the A2A and ACP working groups. The convergence implicitly acknowledges a systemic gap: A2A’s minimalist approach cannot ensure trust propagation in complex agent ecosystems, while ACP’s flexibility undermines reproducible security. The next-generation unified standard must therefore balance \textit{strict security enforcement} with \textit{controlled extensibility}, addressing the very deficiencies in authentication, authorization, and integrity that our results have empirically demonstrated.

\subsection{The Payment Protocol Dilemma: Google's AP2 and CORAL's Validation}
\label{subsec:payment_dilemma}

A second major industry shift is Google's recent announcement of the dedicated Agent Payments Protocol (AP2) \cite{ap2_announce}, designed specifically for agent-initiated financial transactions. This development validates a core premise of our research: agent-based payments cannot be treated as just another generic task. they demand a specialized, high-security architecture.

This is precisely where the architectural design of CORAL stands in stark contrast. Our analysis identified CORAL as the only protocol that integrates a secure payment architecture, through on-chain smart contracts, as a first-class component rather than an afterthought. While Google's approach is to \emph{decouple} payments into a new, purpose-built protocol (AP2), CORAL's approach is to \emph{embed} them securely from the ground up. Both strategies affirm that a simple, general-purpose protocol like A2A is insufficient for high-stakes financial transactions, lending strong empirical and architectural support to CORAL's forward-looking design.

\subsection{The Security Dichotomy of CORAL: Strong Architecture vs. Flawed Implementation}
\label{subsec:coral_dichotomy}

Our empirical validation of the official CORAL Ktor server revealed a pronounced security dichotomy. While the protocol’s architecture demonstrates strong, modern design principles, its current public implementation contains several critical vulnerabilities that compromise these theoretical protections.

\subsubsection{Architectural Strengths (Empirically Validated)}
\label{subsubsec:coral_strengths}

The experimental tests confirmed two major architectural strengths in CORAL’s implementation:

\begin{itemize}
    \item \textbf{Message Integrity (via Transport-Layer Security):}  
    The \emph{Data Tampering} and \emph{Replay Attack} tests, executed through direct HTTP POST requests, were entirely neutralized. In every trial, the server returned a \texttt{400 Bad Request} response (“Transport not found”), indicating the use of a transport-locking mechanism that rejects messages not originating from authenticated transport channels (active SSE connections). This design effectively mitigates a full class of unauthenticated message-forgery attacks and constitutes a major integrity advantage compared to other protocols.

    \item \textbf{Confidentiality (Session Isolation):}  
    The \emph{Potential Excessive Exposure of Data} test, including PII-leakage scenarios, showed perfect session isolation (100\% defense success). Sensitive data sent to Session A was never observable from Session B, confirming robust per-session segregation and strong confidentiality guarantees.
\end{itemize}

\subsubsection{Implementation-Level Vulnerabilities (Empirically Validated)}
\label{subsubsec:coral_vulnerabilities}

Despite its sound architectural design, three critical vulnerabilities were consistently reproducible (100\% success rate) in the official server implementation:

\begin{enumerate}
    \item \textbf{Authentication Bypass (SCA Impersonation):}  
    The SSE connection endpoint (\texttt{/sse/v1/...}) fails to validate the \texttt{privacyKey}. Attackers with a valid \texttt{sessionId} can connect to any session’s SSE stream using an arbitrary or incorrect key. This constitutes a critical authentication flaw (CWE-287).

    \item \textbf{Authorization Bypass (Spoofing):}  
    The SSE endpoint fails to enforce \texttt{agentId} validation. An attacker can connect using a fabricated identifier not present in the session’s Agent Graph. Although the mismatch appears in server logs, the connection remains active, constituting a severe Broken Access Control vulnerability (OWASP A01:2021).

    \item \textbf{Denial of Service (Registry Pollution):}  
    The session-creation endpoint (\texttt{/api/v1/sessions}) lacks rate limiting. During tests, the client successfully created 20 sessions and 20 concurrent SSE connections in rapid succession, proving the presence of an uncontrolled resource consumption vector (CWE-400) enabling denial-of-service conditions.
\end{enumerate}

These implementation-level flaws do not necessarily imply deficiencies in the CORAL \emph{standard} itself. rather, they highlight the security fragility of its current public \emph{implementation}. The vulnerabilities discovered in the SSE subsystem are particularly concerning, as they undermine the very transport-layer integrity mechanisms that form one of CORAL’s strongest architectural defenses.

\subsection{Synthesis of Findings}
\label{subsec:synthesis}

Our cross-protocol analysis reveals a multi-agent communication landscape in transition. The legacy models (A2A and ACP) are being phased out, largely due to the same architectural weaknesses that our empirical validation confirmed, optional security enforcement, weak access control granularity, and the absence of dedicated mechanisms for secure payments.

In contrast, CORAL represents the most forward-looking architectural blueprint by integrating payments and transport-layer security as native design principles. However, our empirical tests show that its current public implementation contains fundamental authentication and authorization flaws within its SSE gateway. This discrepancy illustrates a broader industry pattern: the design of secure, distributed agent systems is progressing faster than the maturity of their reference implementations.

These findings highlight a systemic challenge for the field, bridging the gap between \emph{protocol specification} and \emph{operational security}. They also expose several open challenges that motivate further empirical and standardization work, paving the way for the discussions in the following sections on limitations and future directions.

\section{Limitations}
\label{sec:limitations}

While our empirical findings are robust and reproducible, the present study is subject to several methodological and scope-related limitations that should guide the interpretation of its results.

First, the empirical evaluation of CORAL focused primarily on its official Ktor server implementation, emphasizing off-chain components. We did not perform a formal audit of the Solana-based on-chain smart contracts for vulnerabilities such as re-entrancy or oracle manipulation, which therefore remain theoretical risks (as noted in Section \ref{sec:vulns}). 

Second, the \emph{Message Tampering} test was limited to HTTP POST forgery attempts rather than full man-in-the-middle (MITM) interception of live SSE streams, due to ethical and technical constraints. 

Finally, the ACP testbed, although based on the official SDK, was executed in a controlled environment simulating commonly insecure deployment configurations. Real-world implementations, particularly in large-scale production environments, may exhibit different security characteristics depending on enforcement policies and network conditions.

These limitations do not undermine the validity of the empirical results but rather delineate their scope. The findings should thus be interpreted as conservative lower bounds on the true resilience or vulnerability of each protocol.

\section{Future Work}
\label{sec:future_work}

The rapid evolution of the agent communication landscape, marked by major protocol consolidations and new architectural paradigms, defines several clear directions for future research. The unification of A2A and ACP, alongside Google's introduction of the specialized Agent Payments Protocol (AP2), occurred during the final stages of this study and thus remain outside the scope of our empirical analysis. These developments, however, offer a natural roadmap for subsequent work.

\begin{itemize}
    \item \textbf{Empirical Analysis of Protocol Convergence:} The announced merger of A2A and ACP reflects an industry acknowledgment of the distinct weaknesses identified in both. Future research should conduct a full empirical evaluation of this unified protocol to determine whether it effectively resolves legacy vulnerabilities, particularly in data integrity, authorization, and the enforcement of mandatory security controls.

    \item \textbf{Comparative Study of Payment Architectures:} The emergence of Google's AP2 introduces a new dimension for analysis. A systematic comparison between AP2's decoupled payment model and CORAL's integrated on-chain escrow architecture would yield critical insights into which paradigm, separation or integration, provides stronger guarantees for agent-based financial transactions.

    \item \textbf{Attack-Chaining Dynamics in CORAL:} Our empirical results revealed both strong defenses (transport-layer message validation) and severe weaknesses (SSE authentication bypass). A key next step is to investigate whether these can be combined in "attack chains," where a successful impersonation grants access to a valid \texttt{transportId}, which is then exploited for prompt injection or data tampering. Such analysis would clarify whether CORAL's security strengths can be systematically undermined through multi-stage exploitation.
\end{itemize}

Overall, the next generation of research should focus on evaluating these evolving protocols under realistic threat models, expanding from single-vector analyses toward comprehensive, chained adversarial testing across architectures.

\section{Conclusion}
\label{sec:conclusion}

This study presented the first empirical, comparative security analysis of two active multi-agent communication protocols, CORAL and ACP, benchmarked against a literature-based evaluation of A2A. The results demonstrate a persistent and measurable gap between theoretical protocol design and the actual security of their implementations in real-world environments.

Our experiments uncovered a pronounced security dichotomy within CORAL: its \emph{architecture} is inherently robust, integrating advanced mechanisms such as transport-layer message validation and strong session isolation. Yet, its official \emph{implementation} exhibits critical vulnerabilities, including authentication bypasses (via the \texttt{privacyKey}) and authorization failures (via the \texttt{agentId}) in its SSE gateway, along with exposure to denial-of-service attacks. 

Similarly, the ACP protocol, though architecturally flexible and modular, was empirically shown to suffer predictable weaknesses, specifically, message integrity and confidentiality failures, when its optional JWS validation mechanisms are not strictly enforced. These findings align with ongoing industry developments, such as the A2A/ACP unification, which implicitly acknowledge the architectural shortcomings of earlier designs.

In sum, the current ecosystem of agent communication protocols remains largely insecure. CORAL stands out as the most promising \emph{architectural} foundation, combining integrated payments and secure transport, but its implementation demands immediate remediation. Conversely, A2A and ACP face \emph{architectural} constraints that even correct implementation cannot fully resolve.

Moving forward, this research advocates for a hybrid model that synthesizes CORAL’s integrated payment and transport architecture with ACP’s per-message cryptographic integrity guarantees (e.g., JWS). Such a unified approach, combining secure channels with mandatory message-level verification, offers a concrete pathway toward a resilient, AI-native communication standard suitable for the next generation of autonomous multi-agent systems.

\bibliographystyle{IEEEtran}
\bibliography{protocolscomp}

\end{document}